\documentclass[a4paper,twocolumn]{article}
\usepackage{geometry}
\usepackage{mathtools}
\usepackage[utf8]{inputenc}
\usepackage{authblk}
\usepackage{graphicx}
\usepackage{hyperref}
\usepackage{url}
\usepackage{cite}
\usepackage[normalem]{ulem}

\usepackage{xcolor}
\usepackage{booktabs}
\usepackage{siunitx}
\usepackage{wasysym}
\usepackage[font=small]{caption}
\usepackage{dblfloatfix} 
\usepackage{dcolumn}
\usepackage{bm}

\usepackage{physics}
\usepackage{xr}
\usepackage{listings}
\lstdefinestyle{mystyle}{
  captionpos=b
}
\title{CoolWalks for active mobility in urban street networks}

\author[a,b]{Henrik Wolf}
\author[a]{Ane Rahbek Vierø}
\author[a,c,d,\thanks{}]{Michael Szell}

\affil[a]{NEtwoRks, Data and Society (NERDS), Data Science Section, IT University of Copenhagen, Copenhagen, 2300, Denmark}
\affil[b]{Chair for Network Dynamics, Institute for Theoretical Physics and Center for Advancing Electronics Dresden (cfaed), Technical University of Dresden, Dresden, 01307, Germany}
\affil[c]{ISI Foundation, Turin, 10126, Italy}
\affil[d]{Complexity Science Hub, Vienna, 1030, Austria}

\date{}

\newcommand{\sunsym}{\raisebox{0em}{\scalebox{0.75}{\sun{}}}}
\newcommand{\shadesym}{\raisebox{0.13em}{\scalebox{0.6}{\newmoon{}}}}
\newcommand{\lf}[1][ij]{\ensuremath{\lambda_{#1}}}
\newcommand{\ls}[1][ij]{\ensuremath{l_{#1}}}
\newcommand{\lc}[1][ij]{\ensuremath{l^{\shadesym{}}_{#1}}}
\newcommand{\lh}[1][ij]{\ensuremath{l^{\sunsym{}}_{#1}}}
\newcommand{\Lh}[1]{\ensuremath{L^{\sunsym{},#1,*}}}
\newcommand{\Ltot}[1]{\ensuremath{L^{#1,*}}}

\newcommand{\edge}[2][i]{\ensuremath{#1 \rightarrow #2}}

\newcommand{\graphpath}[3][]{\ensuremath{#2 \rightarrow^{#1} #3}}

\newcommand{\exppathlength}[3][\alpha]{\ensuremath{\Lambda^{#1}_{\graphpath{#2}{#3}}}}
\newcommand{\Path}[3][]{\ensuremath{\Pi^{#1}_{\edge[#2]{#3}}}}
\newcommand{\Cwb}{\mbox{CoolWalkability}}
\newcommand{\Cwbs}{\mbox{CoolWalkabilities}}
\newcommand{\Gcwb}[1][]{\ensuremath{C^\alpha}#1}
\newcommand{\Lcwb}[2][]{\ensuremath{C^\alpha_#1}#2}
\newcommand{\ShadowFrac}[2]{\ensuremath{S_{#1}#2}}

\begin{document}

\twocolumn[
  \begin{@twocolumnfalse}

\maketitle

\begin{abstract}
    \noindent Walking is the most sustainable form of urban mobility, but is compromised by uncomfortable or unhealthy sun exposure, which is an increasing problem due to global warming. Shade from buildings can provide cooling and protection for pedestrians, but the extent of this potential benefit is unknown. Here we explore the potential for shaded walking, using building footprints and street networks from both synthetic and real cities. We introduce a route choice model with a sun avoidance parameter $\alpha$ and define the CoolWalkability metric to measure opportunities for walking in shade. We derive analytically that on a regular grid with constant building heights, CoolWalkability is independent of $\alpha$, and that the grid provides no CoolWalkability benefit for shade-seeking individuals compared to the shortest path. However, variations in street geometry and building heights create such benefits. We further uncover that the potential for shaded routing differs between grid-like and irregular street networks, forms local clusters, and is sensitive to the mapped network geometry. Our research identifies the limitations and potential of shade for cool, active travel, and is a first step towards a rigorous understanding of shade provision for sustainable mobility in cities.
\end{abstract}

{\vskip 10mm}
  \end{@twocolumnfalse}
]

\footnotetext{*Corresponding author. Email: \href{mailto:misz@itu.dk}{misz@itu.dk}}

\section*{Introduction}

To make cities more sustainable and liveable, it is of utmost importance to reduce vehicular traffic and associated car harm \cite{nieuwenhuijsen_implementing_2019,miner_car_2024}, and to promote active mobility like walking and cycling \cite{world_health_organization_regional_office_for_europe_walking_2022}. Active mobility can, for example, be fostered through implementing 15-minute cities, low-traffic neighborhoods, or by improving bicycle infrastructure \cite{nieuwenhuijsen_superblock_2024,szell_growing_2022,brand_climate_2021,steinacker2022,aldred2024impacts,lovelace_propensity_2017}. However, such efforts are only effective if people find the street environment safe and comfortable. An important factor for safety and comfort, apart from low vehicular traffic, is protection from high temperatures and sun exposure. The climate crisis unfortunately comes with global warming and growing temperature variations that create an increasingly hostile outdoor environment \cite{lenton_quantifying_2023}. Cities are especially prone to heat island effects which cause excess deaths from heat exposure \cite{rosenzweig_future_2018}. To combat these effects, \emph{preventing the causes} of the climate crisis should be the top priority \cite{intergovernmental_panel_on_climate_change_climate_2022}. Nevertheless, ensuring acceptable standards of living in cities amid rising global temperatures and increasing urbanization also calls for the urgent exploration of adaptation strategies.

One such adaptation strategy is the provision of shade, to allow pedestrians to avoid time spent in direct sunlight. Shade provision is frequently overlooked in urban planning and climate change mitigation strategies, despite being one of the most efficient and cost-effective ways to reduce heat-related health risks outdoors \cite{turner_shade_2023}. Improving shade provision has the dual benefit of minimizing the harmful impacts of a changing climate while stimulating sustainable modes of mobility that do not contribute to further climate change \cite{litman_cool_2023}. Although there is a growing body of research addressing active mobility \cite{buehler_bikeway_2016,koszowski2019active}, the provision of shade for active mobility is barely explored. In this paper, we start addressing this gap by studying the complex relation between shade distributions and the geometry of buildings and street networks, shedding light on the built environment's effect on shade availability for pedestrians.

The cooling benefits from shade in cities have previously been covered from a variety of perspectives such as urban planning \cite{greenwood2000under, litman_cool_2023}, engineering \cite{rizwan_review_2008, krayenhoff_cooling_2021,li_influence_2020}, or public health \cite{heaviside_urban_2017,lin_shading_2010}. The latter has established that shade provides relief for temperature-related health issues and contributes strongly to a decrease in physiological equivalent temperature \cite{klokAssessmentThermallyComfortable2019, middel50GradesShade2021,turner_shade_2023}, increasing the subjective well-being when traveling.
Furthermore, behavioral data from pedestrian routing in heat has been studied\cite{basu_hot_2024}, and localized mobile phone app prototypes have been developed that implement shade-optimized routing, showcasing the feasibility for solving a practical routing problem for concrete users\cite{olaverri_monreal_shadow_2016, rusig_reducing_2017, deilami_allowing_2020}. These are valuable user-focused case studies that aim for measuring concrete benefits to potential users of an app with a fixed start and end point of a planned trip. 

In particular, Ref.~\cite{olaverri_monreal_shadow_2016} propose and study a pedestrian-tailored navigation application based on path attributes. The study is important pioneering research in this context: It develops a system architecture, database model, routing algorithm, and more, for implementing a mobile phone app with shadow as a tour quality parameter. This app is then evaluated between one particular start and end point in the city of Vienna, comparing sunlight exposure between shaded and classic routing, with valuable conclusions on exposure versus path length. Our contribution takes such first evaluation efforts further and implements a whole mathematical framework for \emph{systematic} investigation, but without an actual app being developed and deployed, to quantify \emph{a city's potential} for walking in shade. Being a city-focused approach, we assess the potential for shaded walking for each point in the city as an aggregate of potential walks from that point to all points in its local neighborhood, going beyond considering raw sunlight exposure on one particular path in one particular city \cite{olaverri_monreal_shadow_2016}. For this task we develop a new metric, the \emph{\Cwb}, with an adequate null model that accounts for best and worst shade coverage in the city, which we systematically evaluate in three different cities. As such, our contribution is more theoretical and methods-focused, coming from an Urban Data Science perspective, in contrast to applied transportation systems engineering approaches \cite{olaverri_monreal_shadow_2016,deilami_allowing_2020}.

Pedestrians are much more exposed to their surrounding environments than motorists \cite{klanjcic_identifying_2022, basu_what_2023}. Most routing for pedestrian mobility thus considers not just travel time or distance, but also, for example, traffic safety \cite{basu_systematic_2022}, land use \cite{basu_what_2023}, the attractiveness of the surroundings \cite{johnson2017beautiful}, or avoiding smells, noises, or other unpleasant sensory experiences \cite{quercia2014a, quercia_smelly_2021}. Previous works on routing have also covered a multitude of computational, environmental, or behavioral aspects of different traffic participants, including traffic assignment \cite{cornacchia2023one}, emissions \cite{cornacchia2022routing,mehrvarz2020optimal}, real-time re-routing and recommendation assistants \cite{falek2022re,pappalardo2024survey}. Like re-routing, considering shade for pedestrian routing is spatiotemporally more intricate than one-time shortest-path routing, since shade availability depends on the interplay between sidewalk network topology and the built environment, together with a dynamic dependence on the time of day.

\begin{figure*}[t!]
    \centering
    \includegraphics[width=0.9\textwidth]{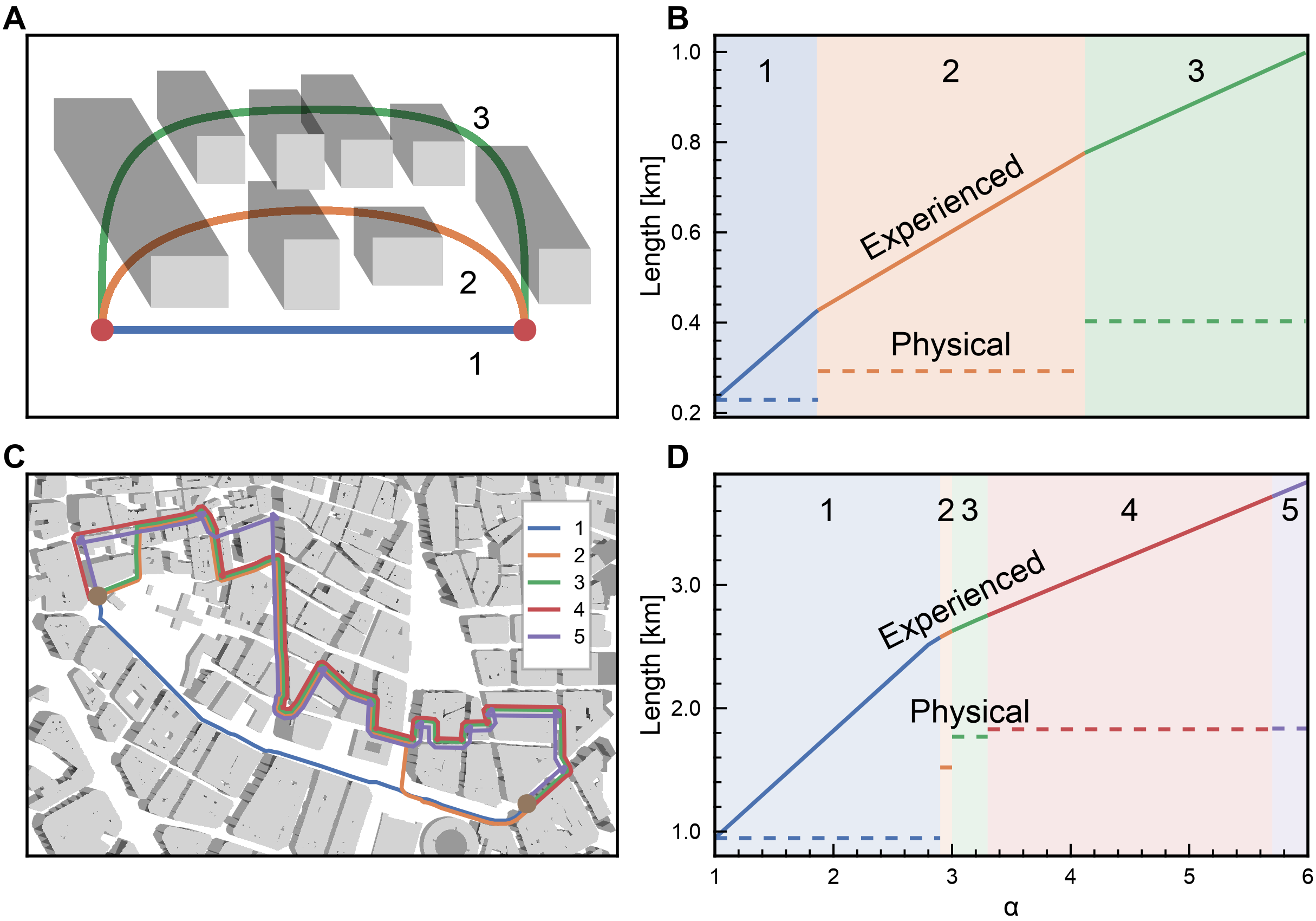}
    \caption{\textbf{Higher sun avoidance $\alpha$ implies choice of routes that
            are not physically shortest but that minimize experienced length.}
        \textbf{A:} Example of three different links connecting two nodes in the street network, from shortest and least shaded (1, blue) to longest and
        most shaded (3, green). \textbf{B:} The choice of which link to take depends
        on the pedestrian's sun avoidance $\alpha$. Increasing sun avoidance
        increases the experienced length because walking in the sun becomes less
        tolerable. At the threshold between different regimes, for example 1 (blue) to 2 (orange),
        the experienced length of the shaded link 2 becomes shorter than the
        experienced length of sunny link 1, implying a preference for link 2 despite
        longer physical length (dashed orange line). \textbf{C,D:} Generalization from links to shortest
        paths with an example of five routes.}
    \label{fig:alpha_changes_paths}
\end{figure*}

To contribute to the fields of active mobility and climate adaptation research, we introduce a route choice model for walking with a sun aversion parameter $\alpha$ and a \Cwb{} metric that operationalizes the potential for shaded routing given an amount and spatial distribution of shade. By studying this metric we reveal the nontrivial impact of street networks and building height distributions on shade provision for active mobility. We measure \Cwb{} for cities with different urban forms and street networks, from grid-like to irregular: Manhattan, Barcelona, Valencia. We introduce diurnal (daily) \Cwb{} profiles and phase portraits to visualize the progression of \Cwb{} in cities over the day. These tools allow us to disentangle the effects of building heights and street geometry and to compare Manhattan's empirical data to the analytical solution for a corresponding regular grid.

We find that the analytic grid solution is independent of an individual's sun aversion $\alpha$, meaning that individuals are not able to minimize time walking in the sun between an origin-destination pair on a perfect grid. However, the slight variations in Manhattan's building heights and street geometry improve the \Cwb{} compared to the exact grid with constant building height, yielding potential benefits for shade-seeking individuals. Moreover, by comparing grid-like cities like Manhattan with more irregular street geometries from Barcelona and Valencia, we uncover different classes of diurnal \Cwb{} profiles that are spatially clustered, meaning that shade availability varies substantially between different parts of the cities. Finally, we discuss the effect of network geometry on the \Cwb{} and compare results between centerline and full pedestrian networks. Understanding how, where, and why the built environment contributes to shade availability are important first steps in adapting cities to a hotter climate and supporting active mobility throughout the year.

\section*{Model and metrics}\label{sec:model}

In this section we first define our route choice model, then the corresponding \Cwb{} metric.

\subsection*{Route choice model with sun aversion $\alpha$}
How a shade-seeking pedestrian selects their routes when traversing a city
depends on the available shade along the
streets. They might choose to walk farther than the shortest path, but with reduced sun exposure, trading a shorter total distance traveled
for a longer distance traveled in the shade. We therefore express the problem of finding the best shaded route in terms of a shortest path problem on the street network $G=(V,E)$ with edge lengths given by the \emph{experienced length}

\begin{equation}
    \lf = \alpha \cdot \lh + \lc \label{eq:experienced_length_edge}
\end{equation}
where \lh{} and \lc{} are the physical lengths in the sun or shade on the street
segment $(i,j)$ connecting intersections $i$ and $j$, respectively. The parameter $\alpha
    \in \left[ 1, \infty \right)$ captures the \emph{sun aversion} of a pedestrian: A constant distance traveled in the sun is experienced as
$\alpha$ times as long as the same distance traveled in shade. An example of the effect of $\alpha$ is shown in Fig.~\ref{fig:alpha_changes_paths}A,B, where the blue link is the shortest but most sun-exposed link preferred by individuals who are not sun-averse (low $\alpha$), while the other links are longer but less sun-exposed preferred by individuals with sun aversion (higher $\alpha$).

Generalizing the experienced length of a link, we define the experienced length of a path $\exppathlength[\alpha]{i}{j}$ as

\begin{equation}
    \exppathlength[\alpha]{i}{j}\left(\Path{i}{j}\right) = \sum_{ab \in \Path{i}{j}} \lf[ab]
\end{equation}
where $\Pi_{\edge{j}}=\left(\edge{k_1}, \edge[k_1]{k_2}, ...,
    \edge[k_{N-1}]{j}\right)$ is a path of length $N$ between $i$ and $j$.

For constant time of day and $\alpha$, we assume that a pedestrian minimizes their experienced path length $\exppathlength[\alpha]{i}{j}$ choosing the path

\begin{equation}
    \Path[*]{i}{j} = \mathrm{argmin}\left(\exppathlength[\alpha]{i}{j}(\Path{i}{j})\right)
\end{equation}
over all possible paths $\left\{\Pi_{\edge{j}}\right\}$ connecting node $i$ to node $j$.

In this context, $\alpha$ is an upper bound on the
maximal increase of the physical length of the shortest path, compared to the
physically shortest path at $\alpha=1$. For example, an $\alpha=1.5$ means that a pedestrian tolerates an increase in path length of up to 50\%. This increase will however only be realized when the shortest path is fully exposed to the sun while the alternative path is fully covered by shade. Another way to think about the value of $\alpha$ is to consider it as the trade-off between physical distance and distance in the sun every individual is willing to endure. Individuals will trade distance in the sun against at most $\alpha$ times as much distance in the shade. If this trade is not realizable, the experienced shortest path will be the same as the physically shortest one.
Figure~\ref{fig:alpha_changes_paths}C,D illustrates how different values of $\alpha$ result in different selected paths through a city.

The exact values of $\alpha$ are not easy to determine empirically, as they might depend on various individual preferences like heat resilience or time constraints, and on local or temporal factors like surface, the height of the sun, or wind conditions. Pedestrians are in general willing to endure detours for a variety of reasons \cite{tong_principles_2022,basu_systematic_2022} compared to the physically shortest path. We therefore set the studied sun aversions to $\alpha \in \left\{ 1.1, 1.25, 1.5,
    2, 4, 10 \right\}$, capturing both realistic values ($\alpha \leq 1.5$) and extreme values ($\alpha \gg 1$) for a comprehensive exploration of the parameter space.

For all shortest-path calculations, we consider fixed times during the
21\textsuperscript{st} of July 2023, not including possible changes in shade
during any trip. This simplification, together with the local weights of experienced edge lengths \lf{},
enables us to efficiently calculate the shortest path for various scenarios using
Dijkstra's shortest path algorithm \cite{dijkstra1959}.

\subsection*{Defining \Cwb{}}

At each point in time $t$, the \emph{shadow fraction} \ShadowFrac{ij}{(t)} of a
street-segment $(i,j)$ is defined as
\begin{equation}
    \ShadowFrac{ij}{(t)} = \frac{\lc{}(t)}{\ls{}},
\end{equation}
denoting the fraction of the segment covered in shade.
Similarly, the \emph{global shadow fraction} \ShadowFrac{}{(t)} denotes the
total shaded length available in the city at time $t$, normalized by the full
length of all streets
\begin{equation}
    \ShadowFrac{}{(t)} = \frac{\sum\limits_{(i,j) \in E} \lc{}(t)}{\sum\limits_{(i,j) \in E} \ls{}}
\end{equation}

\begin{figure}[t!]
    \centering
    \includegraphics{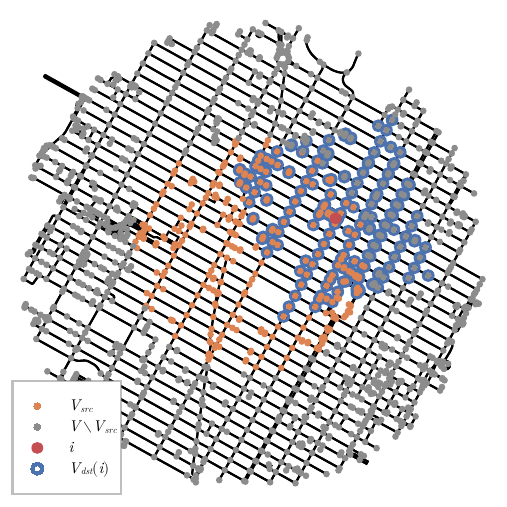}
    \caption{\textbf{Study area definition.} We use a subgraph of the Manhattan street network centered around Times Square; similar subgraphs are extracted from the centers of Barcelona and Valencia. Each considered source node $i \in V_{src}$ (orange) lies within \SI{800}{\metre} of the center and has a set of reachable nodes
        $V_{dst}(i)$, here highlighted in blue for one example node $i$ (red). We limit the number of reachable nodes to a maximum distance of \SI{800}{\metre} on the street network, to effectively capture the local structure of the city. To avoid edge effects caused by cutting the network from the full road network, we include all nodes within \SI{1600}{\metre} of the center. The nodes used in our analysis are therefore the source nodes together with all the nodes in the padding-area (grey).}
    \label{fig:sum_setup}
\end{figure}

This global shadow fraction is a baseline performance measure for a given city which consists of a street network and the associated buildings. We then define the \emph{global \Cwb{}}
\Gcwb[(t)] as
\begin{equation}
    \Gcwb[(t)] = \frac{\sum\limits_{i \in V_{\mathrm{src}}, j \in V_{\mathrm{dst}}(i)}
        \left(\exppathlength[\alpha,*]{i}{j}(0)-\exppathlength[\alpha,*]{i}{j}\left(\left\{\ShadowFrac{ab}{(t)}\right\}\right)\right)}
    {\sum\limits_{i \in V_{\mathrm{src}}, j \in V_{\mathrm{dst}}(i)}\left(\exppathlength[\alpha,*]{i}{j}(0)-\exppathlength[\alpha,*]{i}{j}(1)\right)}
    \label{eq:coolwalkability_definition}
\end{equation}
where $\exppathlength[\alpha,*]{i}{j}$ is the experienced length of the
shortest experienced path between $i$ and $j$, expressed as a function either of the shadow
fraction $\left\{\ShadowFrac{ab}{(t)}\right\}$ at time $t$ or at constant shadow fractions $\ShadowFrac{ij}{}=0$ or $\ShadowFrac{ij}{}=1$ for all edges \mbox{$(i,j) \in E$}, respectively, representing the worst and best case of shade coverage for the city. The \Cwb{}
thus represents how much shorter all experienced trip-lengths are, compared to
the experienced trip lengths in the worst-case scenario with no shade available at all. To make these results comparable between various cities, this value is
normalized by the maximal difference in all experienced trip lengths. Small \Cwb{} values imply lower performance of the city at keeping pedestrians out of the sun, either due to a general lack of available shade or due to the available shade not being distributed in such a way that it can be effectively utilized to reroute pedestrians. Conversely, a \Cwb{} close to one signifies a high performance, due to a generally high availability of shade or a favorable distribution. \Cwb{} is inspired by a recent definition of bikeability \cite{steinacker2022}.

We consider
destinations $V_{\mathrm{dst}}(i)=\left\{j \in V \,\middle|\,\exppathlength[1,*]{i}{j}
    <\SI{800}{\metre}\right\}$ to emulate a limited range
of activity for each pedestrian. This restriction also makes the error negligible from assuming a constant sun position for the duration of a trip, and it is appropriate considering we are interested in the \emph{local} structure of the city at each node and not in potentially long, user-centric trips. Figure
\ref{fig:sum_setup} shows the different parts of the summation setup on the street graph of Manhattan. With an average speed of \SI{5}{\km/\hour}, it would take a pedestrian around
\SI{10}{\min} to complete the longest allowed trip, an interval in which sun position and shade are assumed to not change noticeably. We also implemented a robustness check where we extend the range to \SI{2400}{\metre}, see Supplementary Note 3.

\begin{figure*}[th!]
    \centering
    \includegraphics[width=0.9\textwidth]{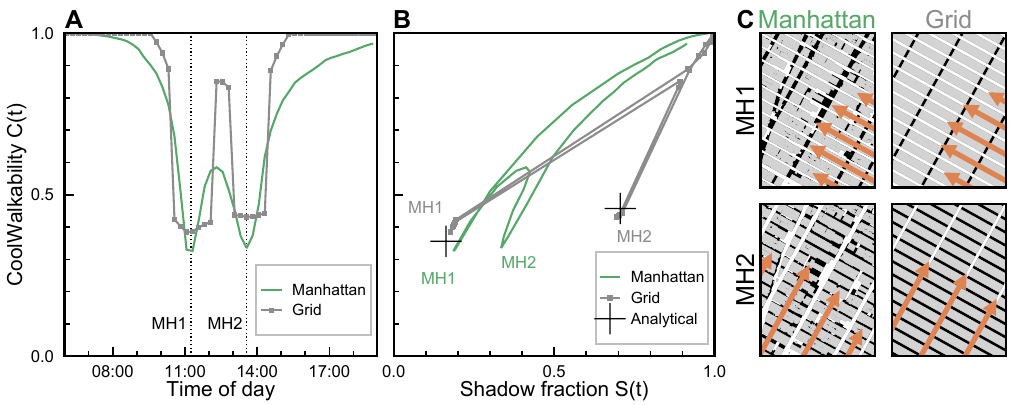}
    \caption{\textbf{Comparing CoolWalkability of Manhattan with a synthetic grid shows similarities but also crucial differences. A:} We introduce the diurnal Coolwalkability profile (here shown for 2023-06-21, and $\alpha=1.5$). It shows two characteristic dips for Manhattan (green) due to the two
        Manhattanhenge events MH1 and MH2 at 11:05 and 13:25, respectively,
        where the sun is aligned with the grid. The dips are also present in
        the synthetic grid (grey), but less pointed. \textbf{B:} We introduce the diurnal CoolWalkability phase portrait. It shows CoolWalkability versus shadow fraction, as functions of time,
        revealing larger differences. Manhattan's portrait (green) is relatively
        smooth due to slight imperfections in its grid structure and
        heterogeneous building heights, while the grid's portrait (grey) jumps
        discontinuously due to its perfect symmetries and constant building
        heights. The
        grid's analytical solution of Manhattanhenges (black crosses) fits
        well with the numerical simulation. \textbf{C:} Illustration of the two 
        Manhattanhenge events MH1 and MH2 for Manhattan and the synthetic grid. Grey
        polygons denote building footprints, black polygons their shadows,
        orange arrows the direction of sun rays.}
    \label{fig:grids_analytical}
\end{figure*}

\section*{Data}\label{sec:data}
In this section we outline the data of street networks and buildings to which we apply our routing model. We focus on subsections from Manhattan, Barcelona, and Valencia. The street networks are taken from OpenStreetMap (OSM) \cite{OpenStreetMap2023}, while the building data were provided by the New York City Office of Technology and Innovation \cite{BuildingsNewYork} for Manhattan and by the General Directorate for Cadastre of Spain \cite{BuildingsSpain} for Barcelona and Valencia.

The sidewalk networks available via OSM can be highly detailed, often including sidewalks on both sides of a street as well as the multiple ways in which pedestrians can cross a street, for example around intersections \cite{rhoads_sidewalk_2023}. We call these high resolution networks the ``full'' pedestrian networks. While these full networks are generally favorable for analyzing pedestrian mobility, OSM data availability depends on the activity of the local OSM communities \cite{haklayHowManyVolunteers2010, neis_street_2012, neis_comparison_2013}. Despite increasing crowd-sourced and remote sensing efforts to collect better data, accurate high-resolution sidewalk networks are currently not available in OSM for most cities around the world \cite{rhoads_sidewalk_2023}. As a useful proxy for a centerline representation of sidewalk networks, where the sidewalks on the two sides of the street are collapsed into a single line in the center \cite{rhoads_sidewalk_2023}, we therefore consider bicycle networks, which we define as all road infrastructure that can legally be used by cyclists. Following this definition, bicycle networks are well mapped in general \cite{ferster2020using, viero_how_2024} because they consist to a large extent of mostly well mapped road networks \cite{barrington-leighWorldUsergeneratedRoad2017}. Bicycle networks come with the additional benefit of small size compared to the full pedestrian networks, with comparable spatial coverage. Further, it is reasonable to assume that the vast majority of the bicycle network can also be reached by pedestrians -- we support this assumption by comparing the bicycle and sidewalk networks for the three study areas, finding a length-overlap of \SI{98}{\percent} (Manhattan), \SI{75}{\percent} (Barcelona), and \SI{80}{\percent} (Valencia). Bicycle networks are thus a reasonable proxy for centerline pedestrian networks in areas that lack a detailed mapping of sidewalks. It also reduces computational complexity, as shown for the study areas in Table \ref{tab:data_overview}.

Our study follows a systematic bottom-up approach, where we start with the coarsest available data or model (centerline network), refining it subsequently with details to understand the impact of such details (building heights, full network geometry, parks, etc.), with the aim to inform future studies on the data or model refinement level necessary to come to reliable conclusions. We thus focus first on the centerline networks, as only they can reasonably be compared to synthetic grid networks and to analytical solutions. Later, in Section \nameref{section:walk_vs_bike}, we compare the differences between centerline and full networks, to assess potential inaccuracies introduced by the centerline approximation. To not miss any such potential inaccuracies, we therefore report results for both centerline and full pedestrian networks for all our analyses whenever possible. As pedestrians do not need to adhere to one-way streets, we ignore edge directions by adding all non-existent reverse edges to the graphs of both network types.

\begin{table}
\begin{center}
    \small
    \begin{tabular}{c|rr|rr|rr}
        \toprule{}
        {} & \multicolumn{2}{c|}{Manhattan} & \multicolumn{2}{c|}{Barcelona}  & \multicolumn{2}{c}{Valencia}\\
        \midrule
        & ctrl                          & full                            & ctrl                          & full        & ctrl      & full\\
        \cmidrule{2-3}
        \cmidrule{4-5}
        \cmidrule{6-7}
        Nodes         & 1506                     & 8702                      & 2780                     & 10975 & 3156 & 8052  \\
        Edges         & 4114                     & 27200                     & 8572                     & 33402 & 9456 & 23858 \\
        Length & 292                      & 845                       & 441                      & 865   & 463  & 693   \\
        Bldgs. & \multicolumn{2}{c|}{6385} & \multicolumn{2}{c|}{14479} & \multicolumn{2}{c}{9483} \\
        \bottomrule
    \end{tabular}

    \caption{Datasets studied in this project. ctrl = centerline network, full = full network. Length in km. \label{tab:data_overview}}
\end{center}
\end{table}

Given the sparse availability of full 3D building data, and for computational simplicity, we handle building data following the 2.5D standard, i.e.~consisting of a footprint-polygon
and a singular height value which is simplified as constant across the whole
building.

In addition to the empirical data, we consider two types of synthetic cities.
On one extreme, we study cities based on an ideal, regular grid with
cell edge lengths $l_a$ and $l_b$. On the other extreme, we construct cities from
the Poisson-Voronoi tessellation \cite{barthelemy_spatial_2022}. In this model, seed-points are first distributed randomly, then Voronoi cells are created around them, and the cell edges represent the city's streets.
For each of these synthetic cities, we either assume a constant height across all buildings, or we sample the building heights from the empirical building height distributions. See Supplementary Note 2 for more details on data generation and processing.

\section*{Results}\label{sec:results}

Before analyzing empirical data from real-world cities, we solve the case of the grid street network analytically, setting the theoretical expectation for \Cwb{} in grid-like cities. We then introduce the diurnal \Cwb{} profile and phase portrait, allowing us to compare deviations of this expectation with a synthetic grid and with the empirical results in Manhattan. This comparison untangles the different factors that cause these deviations: Grid deviations and non-uniform building height distribution. After the grid analysis, we incorporate Barcelona and Valencia, two cities with a more irregular street geometry. Lastly, after observing global differences in \Cwb{}, spatial cluster analysis also reveals \emph{local} differences between grid-like and irregular network structures.

\begin{figure*}[th!]
    \centering
    \includegraphics[width=0.9\textwidth]{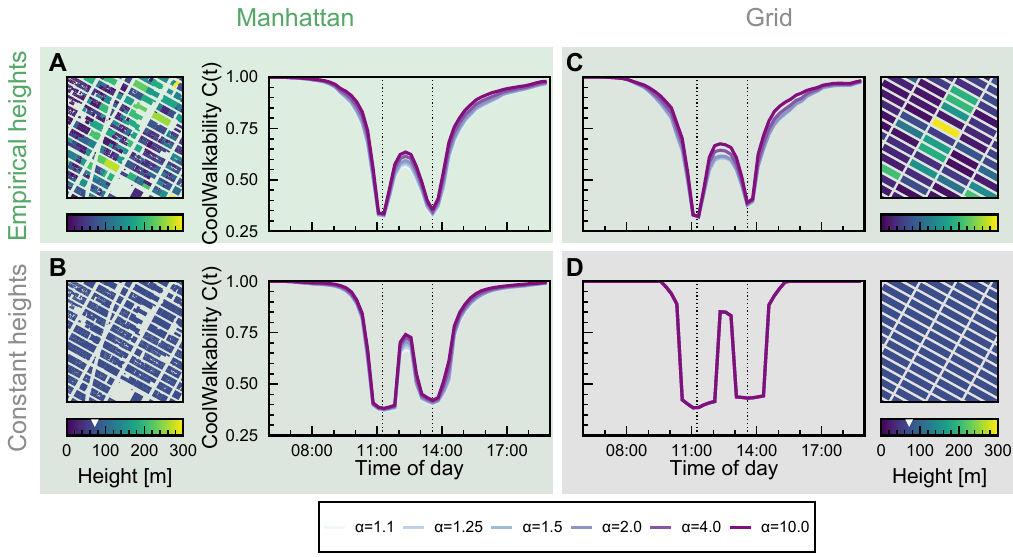}
    \caption{\textbf{Disentangling the effects of building height distribution
            and street geometry on Coolwalkability.} Multiple curves shown in diurnal
        profiles correspond to different sun avoidance values $\alpha$ -- the curves
        mostly overlap, showing independence of Coolwalkability from $\alpha$, as
        proven analytically for the grid. All diurnal profiles shown for 2023-06-21.
        \textbf{A:} Left: Zoom into Manhattan's building footprints. Right: diurnal
        Coolwalkability profile. \textbf{B:} Left: Zoom into Manhattan's building
        footprints at constant height, set as the average \SI{71}{\metre} of
        empirical building heights. Right: The corresponding diurnal
        Coolwalkability profile. Due to loss of building height
        heterogeneity, the Manhattanhenge dips are slightly less pointy.
        \textbf{C:} Right: Zoom into the grid with empirical building footprints taken from
        Manhattan. Left: the corresponding diurnal Coolwalkability profile. Keeping building height heterogeneity but changing from empirical
        street network to grid implies only slight differences in diurnal
        Coolwalkability. \textbf{D:} Right: Zoom into the grid's building footprints at
        constant height, set as the average \SI{71}{\metre} of empirical building
        heights. Left: the corresponding diurnal Coolwalkability profile. Due to loss of building height heterogeneity and the change to a
        grid, the Manhattanhenge dips are considerably broader and less pointy.
    }
    \label{fig:grids_and_perturbations}
\end{figure*}

\subsection*{\Cwb{} on a grid is independent of sun avoidance $\alpha$} \label{section:gridalphaindependence}

A theoretical approximation of a city with a grid-like street network like Manhattan is a regular grid consisting of rectangles with edge lengths $l_a$ and
$l_b$, and buildings of a constant height which have their footprints reduced from the streets by
a distance $w$. On this grid, we derive the CoolWalkability analytically (full derivation in Supplementary Note 1) as a function of the length of shade \lc[a]{} and \lc[b]{} on the
edges:
\begin{equation}
    \Gcwb[{\left(\lc[a], \lc[b]\right)}] = \frac{l_b \cdot\lc[a]+l_a \cdot\lc[b]}{2l_a l_b}
    \label{eq:coolwalkability_analytical}
\end{equation}
This expression is independent of $\alpha$, showing that highly symmetric cities offer pedestrians no choice for more shaded routes, no matter how high their sun aversion. Conversely, the fraction of total street length in shade is given by
\begin{equation}
    S\left(\lc[a], \lc[b]\right)=\frac{\lc[a] + \lc[b]}{l_a+l_b}
    \label{eq:shadow_fraction_analytical}
\end{equation}
which, unlike Eq.~\ref{eq:coolwalkability_analytical}, is symmetric in both
its arguments, showing how the same total \mbox{shadow fraction $S$} may lead to
different CoolWalkabilities, depending not only on the distribution of said
shade but also on the underlying geometry. For example, if $\lc[a]=x$ and \lc[b]=0, we get $S\left(x,0\right) =S\left(0,x\right)$ but $\Gcwb[{\left(x, 0\right)}] = \frac{l_b \cdot x}{2l_a l_b} \neq \frac{l_a \cdot x}{2l_a l_b} = \Gcwb[{\left(0, x\right)}]$ if $l_a \neq l_b$.

\subsection*{Diurnal \Cwb{} profile and phase portrait reveal differences between Manhattan and grid} \label{section:diurnal}
As a tool to visually study \Cwb{} over the course of a day, we introduce the diurnal \Cwb{} profile and phase portrait. Figures~\ref{fig:grids_analytical}A and B show these visualization methods, respectively, comparing the results using empirical data from Manhattan with numerical results in a synthetic city. The synthetic city is based on a regular grid as introduced in the previous section, with parameters chosen to fit the underlying grid structure of Manhattan. Each grid cell has edge lengths of $l_a=\SI{270}{\metre}$ and $l_b=\SI{80}{\metre}$,
and is rotated by \SI{61}{\degree} counterclockwise to align with the Manhattan street grid. The buildings are inset by
$w=\SI{11.5}{\metre}$ and have a constant height of \SI{71}{\metre}, the
area-weighted average of the building heights in the Manhattan dataset.

Comparing the diurnal \Cwb{} profiles, Fig.~\ref{fig:grids_analytical}A, we find a generally similar shape, both showing similar dips around 11:05 and
13:25, caused by the characteristic ``Manhattanhenge'' events where the sun shines
down the urban canyons either from the south-east (MH1) or south-west (MH2)
direction (Fig.~\ref{fig:grids_analytical}C). In both events, the \Cwb{} is roughly similar between the two dips, slightly higher at MH2. In the synthetic grid (grey), the \Cwb{}
shows discontinuities, caused by the high symmetry of the gridded city, where
all shadows sweep over the streets at the same time. These discontinuities disappear for the empirical data (green). We explore this smoothening in the next section.

When comparing the \Cwb{} against the shadow fraction in a diurnal phase portrait, Fig.~\ref{fig:grids_analytical}B, we see more substantial
differences, especially around the second Manhattanhenge event MH2. Here, the shadow
fraction on the grid is considerably larger than the one in Manhattan, without
showing a proportional increase in \Cwb{}. This observation is consistent with
the theoretical predictions from Eqs.~\ref{eq:coolwalkability_analytical} and \ref{eq:shadow_fraction_analytical} because $\Gcwb[{\left(\lc[a], 0\right)}] \approx \Gcwb[{\left(\ls[a], 0\right)}] = \Gcwb[{\left(0, \ls[b]\right)}] \approx \Gcwb[{\left(0, \lc[b]\right)}]$ for $w \ll \ls[a]$ and $w \ll \ls[b]$.

Around MH2, Manhattan has much less shade available
compared to the grid but is still able to provide about as good a \Cwb{} as the
grid. In general, the empirical curve (green) runs to the left or above the grid curve (grey) during corresponding times of the day, implying a better potential use of shade for walking than expected for a perfect grid with constant building heights. This situation is different for Barcelona and Valencia, which are less grid-like and have lower, more uniform building heights, therefore not featuring distinct ``henge'' events, see Supplementary Figures~SI1 and SI2.

\subsection*{Empirical building heights smoothen diurnal profiles} \label{section:untanglingeffects}
To untangle the different effects leading to deviations between empirical and synthetic data, as just observed in Fig.~\ref{fig:grids_analytical}B, we investigate the impact of a constant building height distribution,
by running the same study on Manhattan but with constant building heights
of \SI{71}{\metre}, and on the grid with building heights randomly drawn from the
area-weighted distribution observed in Manhattan. The results of this experiment
are shown in Fig.~\ref{fig:grids_and_perturbations}. Going from the fully
empirical Manhattan (Fig.~\ref{fig:grids_and_perturbations}A) to the case of Manhattan with constant building heights (Fig.~\ref{fig:grids_and_perturbations}B), we
find the dips around the two Manhattanhenges to widen and to increase their \Cwb{}
values slightly. Also, the flanks dropping in to MH1 and leading out of MH2 steepen. Going from Manhattan with constant building heights (Fig.~\ref{fig:grids_and_perturbations}B) to the most synthetic case of a grid with constant building heights (Fig.~\ref{fig:grids_and_perturbations}D) iterates the same effect changes: The dips plateau and widen, discontinuities and slopes increase. The effect of going from the fully empirical Manhattan (Fig.~\ref{fig:grids_and_perturbations}A) to the grid with empirical heights (Fig.~\ref{fig:grids_and_perturbations}C) is mostly a slight increase in \Cwb{} values, but keeping a similar pointedness of Manhattanhenge dips. The largest effect is observed in the grid case, going from empirical to constant heights (Fig.~\ref{fig:grids_and_perturbations}C to D). The two dips go from sharp to two plateaus stretched over around 1.5 hours each.

To summarize, going from empirical to constant building heights has a stronger impact on diurnal \Cwb{} profiles than going from Manhattan to the perfect grid. Manhattan's grid imperfections have a negligible influence on \Cwb{}. An implied consequence for future modeling efforts is to prefer approximating a grid-like network with a perfect grid than to neglect the heterogeneous distribution of building heights. For all these cases, we also observe little to no \mbox{$\alpha$-dependence} of the profiles (they mostly overlap), further showing how
there are nearly no relevant choices to be made by pedestrians in such grid-like
networks.

\subsection*{\Cwb{} clusters locally}

\begin{figure*}[t]
    \centering
    \includegraphics[width=0.7\textwidth]{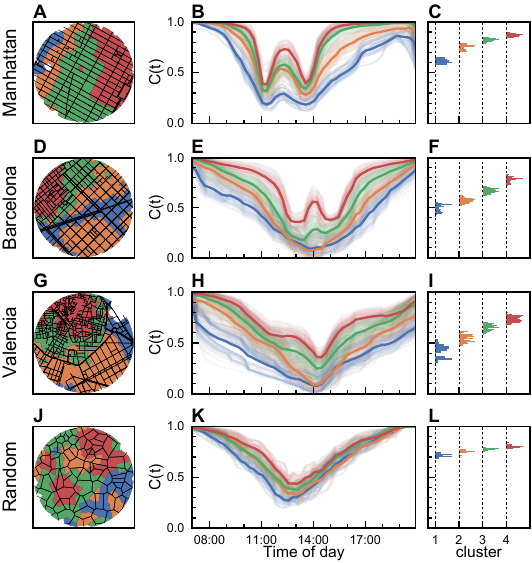}
    \caption{\textbf{Spatial clustering by CoolWalkability leads to areas with different profiles.} From top to bottom, we study the cities Manhattan, Barcelona, Valencia, and the random null model (Poisson-Voronoi). Left column: Clustering local Coolwalkability of each node in the street network leads to spatial clusters of similar CoolWalk potential. Middle column: The diurnal profiles of these clusters display high variations within each city and between different cities. In particular, the more organic, least grid-like areas (red curves) display the highest potential. \textbf{K:} The null model shows the baseline of small variation. Right column: the distributions of the time average of each diurnal profile within each cluster illustrate the large potential differences in empirical street networks. \textbf{L:} These differences are negligible in the null model.}
    \label{fig:clustering_reveals_spatial_order}
\end{figure*}

The global \Cwb{} is aggregated over the whole study area, neglecting
spatial heterogeneities within cities that might affect \Cwb{} locally. To study
the impact of the geometric variations in building and street geometry, we
calculate the \emph{local \Cwb} \Lcwb[i]{(t)} for each vertex $i \in V_{src}$
as
\begin{equation}
    \Lcwb[i]{(t)} = \frac{\sum\limits_{j \in V_{\mathrm{dst}}(i)}
        \left(\exppathlength[\alpha,*]{i}{j}(0)-\exppathlength[\alpha,*]{i}{j}\left(\left\{\ShadowFrac{ab}{(t)}\right\}\right)\right)}
    {\sum\limits_{j \in V_{\mathrm{dst}}(i)}\left(\exppathlength[\alpha,*]{i}{j}(0)-\exppathlength[\alpha,*]{i}{j}(1)\right)}
    \label{eq:coolwalkability_local}
\end{equation}
for multiple times across
the whole day, constantly spaced \SI{15}{\minute} apart. Interpreting the
resulting discrete series $\Lcwb[i]{(t_j)}$ as an element of a $j$-dimensional
vector space, we map the problem of
distinguishing qualitatively different temporal trajectories to a clustering
problem, which we solve using a combination of DBSCAN and $k$-means: In the first step, we detect and eliminate outliers using DBSCAN and find clusters within the remaining intersections using $k$-means. The resulting clusters represent vertices in the street network that show similar diurnal \Cwb{} profiles.

Visualizing the spatial relationship between the vertices in real cities reveals spatially coherent structures (Fig.~\ref{fig:clustering_reveals_spatial_order}A,D,G). Cities have nontrivial spatial
variations of their urban form which are picked up by
our clustering approach. This variation is reflected in the noticeably different local \Cwbs{} between the clusters (Fig.~\ref{fig:clustering_reveals_spatial_order}B,E,H).

For both Barcelona (Fig.~\ref{fig:clustering_reveals_spatial_order}D) and Valencia (Fig.~\ref{fig:clustering_reveals_spatial_order}G), the clusters are well explained by the geometric heterogeneity in the street layout. In Barcelona, the cluster with the lowest average local \Cwb{} (blue) captures the vicinity of the Avinguda Diagonal, Barcelona's widest street. The remaining clusters align well with the three districts which make up the study area: Eixample  (orange), Gracia Nova (green), and Gracia (red), in order of increasing average \Cwb{}. Interestingly, the profiles of the orange and green clusters do not exhibit the double-well structure we found earlier for the regular, grid-like network structure of Manhattan. In these two areas of the city, the low height of the buildings combined with the large width of the streets and the high angle of the sun causes the buildings to not cast any shade on the street network geometries which are generally located in the middle of the street. As such, the grid structure is not apparent during the day. Only in Gracia (red), where the streets are generally narrower, do we find a remnant of this effect.

We also find substantially different clusters in Manhattan (Fig.~\ref{fig:clustering_reveals_spatial_order}A,B,C) although Manhattan's street network is highly symmetric and thus cannot be expected to explain the structure of the obtained clusters. However, the spatial distribution of building heights can. All clusters show the
characteristic double well (Fig.~\ref{fig:clustering_reveals_spatial_order}B), albeit with different strengths, increasing from west to east. The building height in and around the blue cluster is lower than for the other clusters. The last, red cluster receives the most shade because it has the highest buildings. Of special note are the asymmetries between the decrease and increase of \Cwb{} in the blue and orange clusters. While the \Cwb{} of the green and red clusters decreases and recovers quickly in the two Manhattan-Henge events, the recovery is noticeably slower for the blue and orange clusters. In the morning these regions are shaded mostly by the higher buildings
east of them, but are less shaded by the comparably lower buildings in the
west. As the sun sets, the shorter shadows grow longer, causing the \Cwb{} to recover, but slower. Similar temporal asymmetries can thus be expected in any city where there is a non-zero gradient in the spatial building-height distribution.

For the synthetic city based on the Voronoi tessellation (Fig.~\ref{fig:clustering_reveals_spatial_order}J,K,L) we also observe patches, but spatially less coherent. The low but non-zero coherence is explained by the limited radius of movement imposed when calculating
\Lcwb[i]{(t)}, as two physically close nodes $i$ and $j$ tend to be
close in network distance as well. Therefore, the intersection of their
respective reachable destinations $V_{dst}(i) \cap V_{dst}(j)$ is large, and the
sums in Eq.~\ref{eq:coolwalkability_local} will produce similar results.
In other words, places close to one another have similar CoolWalkabilities \emph{by construction}. This locality principle is the case for \emph{any} network, but the question is how distinct the spatial clusters in empirical networks are from each other compared to clusters in the random Voronoi model. Indeed, the qualitative differences between the \Cwb{} profiles of different clusters in the Voronoi model (Fig.~\ref{fig:clustering_reveals_spatial_order}K,L) is much smaller than for the real cities, which is especially apparent in the \Cwb{} variation plot (Fig.~\ref{fig:clustering_reveals_spatial_order}L). This discrepancy provides evidence that irregular networks, narrower streets, and/or heterogeneous building height distributions of real cities have a non-trivial impact on local \Cwb{}. The same figure but for constant building heights shows similar behavior, and even smaller \Cwb{} variations for the clusters in the Voronoi model (Fig.~SI3L). 

Whether streets need to be narrower (or buildings higher) to provide better local \Cwb{} remains a question for future research \cite{turner_shade_2023}.

\begin{figure*}[t]
    \centering
    \includegraphics[width=\textwidth]{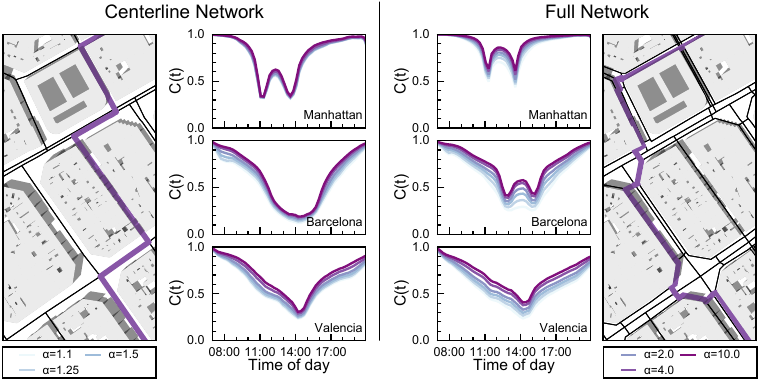}
    \caption{\textbf{Network geometry can have an impact on \Cwb{}.} Considering centerline versus the full pedestrian networks can lead to different results, as links in these networks are sometimes mapped differently in OpenStreetMap. Left map: A centerline-based mapping in Barcelona can inaccurately reflect the actual position of sidewalks. Right map: The full pedestrian links in the same area are mapped with more detailed geometries, some of them closer to buildings and therefore shaded, which comes with more flexibility for shaded routing when walking. Middle: Accordingly, the diurnal profiles in the centerline and full networks can differ substantially, especially for high sun avoidance $\alpha$: Full networks generally display higher \Cwb{} and are more likely to have two dips.}
    \label{fig:network_geometry_matters}
\end{figure*}

\subsection*{Network geometry matters} \label{section:walk_vs_bike}
By ``network geometry'' we understand the spatial geometry used to represent the physical position of elements in a transport network. To which extent this geometry is an accurate reflection of reality depends on many factors and choices, such as the data model, data resolution, or mapping practices, which all can depend on the use case (like routing) and on the transport mode \cite{rhoads_sidewalk_2023,haklayHowManyVolunteers2010,viero_how_2024}.

In our data source OpenStreetMap, and many other road network datasets, the mapped geometry that represents a street is the street's centerline \cite{rhoads_sidewalk_2023}. Although pedestrians usually do not get space allocated on the centerline, but on the side of the street, centerlines are
generally considered sufficient for deriving meaningful quantities like the length of a street segment. This proxy is usually adequate enough to enable effective research and applications on street networks, including routing for cyclists
or pedestrians \cite{steinacker2022,rhoads_sidewalk_2023}. However, in our specific use case, the centerline is not necessarily a reliable proxy, as the quantities of
length in shade \lc[a] and length in the sun \lh[a] on all edges $a$ can
depend on the exact location of the geometry of each sidewalk in
relation to the surrounding buildings. This dependence is especially relevant if the length of the shadows is in the order of the street width, like in Barcelona.

Solving this accuracy problem requires either to offset the location of the centerline geometries towards the sidewalks where pedestrians actually walk, or to acquire higher resolution networks which include the sidewalks as their own geometry. While the first approach would allow for more widespread application even for cities for which high-resolution sidewalk geometries are not available, this requires additional information on
the width of each street, which is often not available. This lack of exact geometry and width information is a known data limitation in sidewalk networks \cite{rhoads_sidewalk_2023}. Fortunately, the OSM dataset of Manhattan, Barcelona and Valencia also contains many
sidewalks as separate geometries, enabling the comparison between the centerline-based proxy network and the full, high-resolution pedestrian network.

We therefore calculate the global \Cwb{} as a function of time for various values of the sun aversion $\alpha$ in all three studied cities on both these networks -- centerline and full. The resulting
profiles can show substantial differences,
depending on the observed cities, as shown in Fig.~\ref{fig:network_geometry_matters}. Compared to the centerline networks, in full pedestrian networks the observed \Cwb{} increases for all cities, as the sidewalks are generally closer to the buildings, increasing the probability of one side being shaded. Thus, crossing the street often causes only a small detour while substantially increasing the distance traveled in shade, increasing the
overall \Cwb{} even for small sun-aversions.

In addition to the overall increase in \Cwb{}, we observe an increased sensitivity with the sun-aversion $\alpha$. Where pedestrians on Manhattan's centerline network did not have many options of similar length from which to choose an optimal route given their individual preferences, the full sidewalk network provides a multitude of such options due to the denser network and slight variations in geometry placement of the footpaths. Here the temporal location of the minima does not change, as the sun still aligns with the urban canyons at the same time; the dips do, however, become narrower since the duration for which both sides of the street are exposed at the same time decreases. If only one side is exposed, a pedestrian can often get into the shade by crossing the road and thus increasing the \Cwb{}, especially around the henge events. For Barcelona, the effect of the full pedestrian network is especially obvious as the profiles of the global \Cwb{} change qualitatively to now exhibit a similar double-well structure as Manhattan.

Depending on the observed city, a switch from the centerline network to the full sidewalk network with more accurately mapped pedestrian infrastructure can have either a small impact, as seen for Valencia, a mostly quantitative effect as in Manhattan or even cause a qualitative change in the dynamics, as demonstrated for Barcelona. The same effects apply to the spatial clustering plot (Fig.~\ref{fig:clustering_reveals_spatial_order} vs.~Fig.~SI4), but without changing the main result of \Cwb{} clustering.

\subsection*{Incorporating green spaces} \label{section:parks}
Although buildings are the largest shade-providing objects in cities, urban greenery such as parks and trees should also be considered when studying urban shading, as it is a widely used and flexible approach for enhancing pedestrian comfort in cities. Due to availability and quality issues with open tree data \cite{battiston_need_2024,croeser_acute_2024}, here we use park data as a proxy as they are readily available on OSM (for the studied cities). We used this data to re-run our simulations, modeling each park in our study areas as a canopy of \SI{8}{\metre} height, assuming a park to be fully covered and shaded by trees and able to cast a shadow on surrounding streets. The results are reported in Figs.~SI7-9. This addition increases the CoolWalkability slightly in the full networks, especially around noon. Buildings only cast shade on the streets next to them when the sun angle is sufficiently low, while trees are able to cast usable shade onto paths beneath them even when the sun is at its highest, causing them to have a stronger effect during noon. In centerline networks at same routing length, the effect is overall weaker. 

While there is a measurable effect of parks on the \Cwb{} up to a difference of 0.1, this difference is relatively small and does not qualitatively change any of the \Cwb{} profiles. Thus, this additional analysis shows that there are some local \Cwb{} gains from incorporating parks, but also that an overall assessment of \Cwb{} is robust to neglecting parks, at least in the considered cities.

\subsection*{Realized routes and impacts} \label{section:impacts}
\Cwb{} gives a metric for the performance of a city to provide shaded routing but lacks practical information for individual pedestrians. The chosen $\alpha$ stands for how much farther a pedestrian is willing to walk in shadow than in sun -- but how much longer are such sun-avoiding walks \emph{when actually realized}, and how does this translate into practical, human-readable benefits?

To answer these questions, we introduce two simple measures: First, the relative distance traveled in the sun when optimizing for shaded routes, compared to the distance traveled in the sun when using the physically shortest paths:
\begin{equation}
\frac{\Lh{\alpha}}{\Lh{1}}
\end{equation}
where $\Lh{\alpha}$ is the total length traveled in the sun with a sun aversion of $\alpha$. Second, the relative increase in physical trip-length when optimizing for shaded routes compared to the physical length of the physically shortest paths:
\begin{equation}
\frac{\Ltot{\alpha}}{\Ltot{1}}
\end{equation}
where $\Ltot{\alpha}$ is the total length traveled in the sun with a sun aversion of $\alpha$. 

We discuss in detail the daily evolution of these two metrics in Supplementary Note 4, visualized in panels E-H and in panels I-L of Figs.~SI10-12, respectively. In summary, the first main result is that the overall physical length of all paths increases only by up to \SI{10}{\percent} for $\alpha\leq2$, and only up to \SI{30}{\percent} for $\alpha=10$. These values are thus all considerably lower than the theoretically possible increase by the factor $\alpha$, since in practice physically shortest paths are usually not fully covered in sun with experienced shortest paths being fully covered in shade. Second, the relative decrease of distance in the sun fluctuates mostly between \SI{20}{\percent} and \SI{90}{\percent}, strongly depending on the city, time of day, and sun aversion $\alpha$. In general, we find that the daily curves of $\frac{\Lh{\alpha}}{\Lh{1}}$ and $\frac{\Ltot{\alpha}}{\Ltot{1}}$ are non-trivial and hard to interpret, possibly due to their numerical sensitivity: For example, they are naturally sensitive at extreme values of shade availability, such as during full shade in the morning/evening and little shade at noon. This numerical sensitivity provides another good motivation for our \Cwb{} metric, as \Cwb{} does not suffer from such fluctuations and gives a clear picture of a city's performance for providing shaded walks. 

Despite the difficulty in finding general patterns for these metrics, for a user-centric ``navigation app'' scenario it would be useful for an individual pedestrian to be informed about their length decrease in the sun $\frac{\Lh{\alpha}}{\Lh{1}}$ while choosing $\alpha$, including for medical reasons, as it can be assumed to be proportional to the reduction of heat-stress or UV exposure. Similarly, it would be crucial to be informed about the \emph{actual} total length increase $\frac{\Ltot{\alpha}}{\Ltot{1}}$. Both these values could also be expressed as absolute rather than relative length differences.

Finally, we assessed the number of intersections and left turns encountered as a function of total physical distance, for individual $\alpha$ values, Figs.~SI13-14, as these metrics can serve as a useful proxy for the ``complexity'', effort, or cognitive load associated with navigating a route. We found that the number of intersections encountered tends to be roughly linear in the total physical distance (but we cannot exclude non-linear exceptions for individual $\alpha$ values in cities where long paths could be routed through neighborhoods with distinctly higher intersection densities), and the fraction of left turns is always very close to $\frac{1}{2}$. This result shows that shade-aware routing tends to not increase the complexity of the experienced shortest paths beyond the complexity added due to an increased length.

\section*{Discussion}\label{sec:discussion}

This research represents a first exploration and method development towards the systematic study of shaded routing and the potential benefits that the built infrastructure can provide. We were able to analytically derive solutions for regular grids, to disentangle building height distribution from street network geometry, and we found a particular dependence of results on network geometry, showing higher potential \Cwbs{} for walking than for cycling networks. Below, we discuss the implications of these results, limitations, and relevance for urban planning and future research.

Although the street layout and building heights of existing cities are mostly static and only change slowly over time, our results reveal several insights which might be of relevance for achieving optimal shading in future urban development projects. More pointwise shading is, of course, achieved by narrower streets, taller buildings, or by adding trees or other objects that can block direct sunlight. However, as our analysis considered full paths and not just pointwise shade, it also exposed the nontrivial dependence of \Cwb{} on building height \emph{distributions}: Especially the diurnal \Cwb{} phase portrait (Fig.~\ref{fig:grids_analytical}B) reveals that at a given shadow fraction, more \Cwb{} can be achieved than expected for a perfect grid with constant building heights. We furthermore showed that an irregular street network can provide a qualitatively different diurnal \Cwb{} profile locally (Fig.~\ref{fig:clustering_reveals_spatial_order}), for example by avoiding ``henge'' events and corresponding dips, with very low shade availability.

A large-scale, systematic study of geometric factors such as street bearing entropy \cite{boeingUrbanSpatialOrder2019} in relation to \Cwb{} remains open, which could answer \emph{which} network properties are most beneficial when seeking more shade for active travel. To better understand local contexts and the cause of temporal asymmetries, \Cwb{} could, for example, be investigated on rotated cities, for varying latitudes, or for larger study areas and a larger number of cities. Further, more realistic pedestrian routing scenarios could be considered beyond the sidewalk network perspective, such as traversing open areas \cite{andreev_towards_2015}, and to account for the potentially considerable time or risk involved in crossing a heavily trafficked street. Such an approach is unfortunately limited by the spatial data structure, which in OSM is mostly one-dimensional and thus fundamentally not adaptable to 2D routing \cite{rhoads_sidewalk_2023}. Further, OSM usually lacks information whether an edge represents a crossing within a street intersection or between two ``regular'' sidewalks across the two sides of a street, thus making it practically impossible to study street crossings.

Shade optimization is also not only a question of sun avoidance during hot seasons, because in other seasons too much shade can be undesirable -- as demonstrated, for example, in a 10-year long study in Taiwan \cite{lin_shading_2010}. To identify the right annual balance between shade and sun provision depending on the local context, our model could easily be extended to also account for sun \emph{seeking}. Further, the assumed form of experienced length $\lambda_{ij}$ used in Eq.~\ref{eq:experienced_length_edge} is a choice that facilitates analytical convenience and reflects reality in a first approximation, but could be extended. Fundamentally, we do not know $\alpha$ (or its distribution among people and contexts), which would need to be measured and calibrated, for example through field experiments. The dependence could also be non-linear, for example to preferentially avoid \emph{continuous} gaps in the sun, or to account for vector or visibility based pedestrian navigation \cite{bongiorno2021vector}.

In order to make the methods presented in this paper of further relevance to urban planning applications, researchers need access to data for pedestrian and cyclist routing of a much higher quality. The marginalization of active mobility like walking and cycling does not only happen on the policy level, but also on the level of data quality and completeness \cite{viero_how_2024,ferster2020using,rhoads_sidewalk_2023,zielstra_using_2012}. Specifically, the lack of detailed, topologically correct, and spatially accurate data both on sidewalk and bicycle networks means that results on whether a given edge will be in shade or sun often come with substantial uncertainties. Similarly, differing mapping practices around intersections, a lack of common mapping and quality standards for street network data from the perspective of active mobility, and uncertainties on how network simplification can influence results warrant further work on active mobility networks \cite{boeing2017osmnx, pung2022road, rhoads_sidewalk_2023, zielstra_using_2012}. In addition to data quality issues for street network data, building height data is only available locally \cite{biljeckiQualityCrowdsourcedGeospatial2023a} and often follow different standards \cite{milojevic-dupontEUBUCCOV0European2023}. However, new GeoAI and image recognition tools are showing promise towards creating global, high-resolution data sets \cite{biljecki2015applications,biljecki2021street,hamim2023mapping,treccani2023automating}.

Our research on the status quo has just started to scratch the surface of a much-neglected topic in urban planning \cite{turner_shade_2023}. Besides describing the existing situation, future research should ask how to use the know-how on \Cwb{} to effectively design better cities. How to identify the most optimal improvements to \Cwb{} within a given budget? The non-trivial impact of buildings, which both radiate heat and provide shade \cite{rizwan_review_2008,li_influence_2020}, highlights that our focus on buildings and street networks is limited and part of a broader interdisciplinary puzzle \cite{nieuwenhuijsen_heat_2019,koszowski2019active}, which should be extended with further work on potential urban interventions like tree planting or installation of (solar panel) canopies \cite{turner_shade_2023}. We considered as a first step buildings only and not trees, due to availability issues with tree data. Buildings are moreover the largest shade-providing objects in cities, but are often ignored in the shade provision literature compared to trees \cite{turner_shade_2023}. Shade should ideally become a fundamental part of infrastructure planning via ``shade master plans'' and other policy guidelines \cite{turner_shade_2023,litman_cool_2023}, and these plans should be based on evidence that research like ours can provide. 

Central for defining the ``optimality'' of improvements should be the notion of equity \cite{pereira2017distributive}. Previous research has shown that urban ``shade deserts'' -- places lacking the shade needed to reduce heat burden and protect human health outdoors -- are part of the lived experience for low-income communities, and exacerbate heat-related health disparities \cite{turner_shade_2023}. Such marginalized communities have less access to shade and to air conditioning, and it is thus crucial to consider how shade provision can target those most in need. 

Although our paper focuses on the health aspects of shaded mobility, further practical considerations could include economic aspects such as changing foot traffic to businesses. For example, the prospect of more shaded routes at tall buildings could divert people away from open spaces. If there was a spatial correlation between such aspects of urban form with local businesses, such businesses could either be harmed unintentionally, or they could benefit. In the long term, following the increased impacts of climate change, businesses will aim to relocate to shaded areas, or they will have increased interest in more shade provision, which makes our study also relevant in this context. Future work should thus incorporate data of human mobility and local businesses \cite{yabe2024behavior}, and extend models of urban business location \cite{stahl1987theories}  to account for shade provision and economic consequences.

Finally, despite the multidimensional benefits of better shade provision -- including more vegetation or green canopies -- on liveable cities, in the context of climate change, it must be stressed that these are primarily adaptation strategies and thus do not address the underlying core issue of human-caused increases in temperatures and extreme weather events \cite{intergovernmental_panel_on_climate_change_climate_2022}.

\section*{Data availability}
All data used for the experiments are publicly available at \url{https://doi.org/10.5281/zenodo.11103149} \cite{dataset}.

\section*{Code availability}
The open-source Julia code repository is available at \url{https://github.com/henrik-wolf/CoolWalks}.

\section*{Author contributions}
H.W.~developed the code, conducted the experiments, performed the analytical derivations, and produced all data visualizations. All authors contributed to conceiving the model, the experiments, and visualizations. All authors analyzed the results, wrote and reviewed the manuscript. M.S.~led the project and the writing. All authors have read and agreed to the final version of the manuscript.

\section*{Competing financial interests}
The authors declare no competing interests.

\section*{Acknowledgments}
We thank Roberta Sinatra for brainstorming the idea for this project, and Marc Timme and Malte Schröder for helpful discussions. M.Sz.~acknowledges funding from EU Horizon Project JUST STREETS (Grant agreement ID: 101104240). All network data from OpenStreetMap \cite{OpenStreetMap2023}.

\end{document}


\maketitle

\noindent This is the supplementary information for the manuscript containing supplementary notes and figures.\\[1cm]

\setcounter{page}{1}
\setcounter{figure}{0}

\section*{Supplementary Note 1: Data acquisition}\label{appendix:datasets}
\subsection*{Street networks} \label{street_network_queries}
We used OpenStreetMaps as the source for the studied street networks of Manhattan, Barcelona and Valencia, which can be conveniently accessed via the Overpass query language. Below are the query strings we used for downloading the pedestrian and cycling networks, respectively. Note the use of \texttt{[bbox:\{\{bbox\}\}]} as a spatial filter, used here to enable copying into overpass turbo. In the real application, this was replaced by the bounding box of the study area.

\begin{lstlisting}[caption=Overpass query for pedestrian networks.]
[out:json][timeout:180][bbox:{{bbox}}];
way["highway"]["area"!~"yes"]
  ["highway"!~"cycleway|motor|proposed"]
  ["highway"!~"construction|abandoned|platform|raceway"]
  ["foot"!~"no"]
  ["access"!~"private"]
  ["service"!~"private"];
(._;>;);
out count;
out ;
\end{lstlisting}

\begin{lstlisting}[caption=Overpass query for cycling networks.]
[out:json][timeout:180][bbox:{{bbox}}];
way["highway"]
  ["area"!~"yes"]
  ["highway"!~"footway|steps|corridor|elevator|escalator"]
  ["highway"!~"motor|proposed|construction"]
  ["highway"!~"abandoned|platform|raceway"]
  ["bicycle"!~"no"]
  ["access"!~"private"]
  ["service"!~"private"];
(._;>;);
out count;
out ;
\end{lstlisting}

When loading the networks from the resulting files, we ignore possible one-way streets by adding all non-existent reverse edges to the street graph.

\subsection*{Buildings}
\subsubsection*{New York City}
The building data for Manhattan was provided by the New York City Office of Technology and Innovation \cite{BuildingsNewYork}. It includes the footprints of New York City buildings as well as the height of their roofs above the ground.

\subsubsection*{Spain}
The building data for Barcelona as well as Valencia was provided by the General Directorate for Cadastre of Spain \cite{BuildingsSpain}. The available datasets contain the building footprints for the buildings in the respective region together with additional polygons representing segments of the buildings which can be used to map variations in height across the associated building. Neither the footprints of buildings nor their segments come with a directly accessible height value, and only the segments have a value for the number of floors available. We therefore assume a constant height of \SI{4}{\metre} per floor.

It would, in theory, be possible to use the segments transformed in this way directly as the input for our analysis. However, the large amount of segments compared to the number of buildings would result in increased computational requirements. As such, we approximate the height of each building by the area-weighted average of the heights of each of its segments.

\section*{Supplementary Note 2: Analytical derivations}\label{appendix:derivations}
\subsection*{CoolWalkability on grids - general form}
Starting from main text Eq.~6,
we derive the
analytical expression for the CoolWalkability on a city with a perfect, infinite grid-like
street network and buildings of constant height and shape.\\
When writing main text Eq.~1
 for $\lh{}=0$
\begin{equation}
    \lf{} = \lc{} = \ls{}
\end{equation}
and $\lc{}=0$
\begin{equation}
    \lf{} = \alpha \cdot \lh{} = \alpha \cdot \ls{}
\end{equation}
we see that for these extreme cases, the felt lengths only differ by the
constant factor $\alpha$. As such, paths which minimize $\exppathlength[\alpha]{i}{j}(0)$ minimize
$\exppathlength[\alpha]{i}{j}(1)$ as well. More specifically, we get $\exppathlength[\alpha,*]{i}{j}(0) = \alpha
    \cdot \exppathlength[1,*]{i}{j}$ and $\exppathlength[\alpha,*]{i}{j}(1) = \exppathlength[1,*]{i}{j}$. Plugging these results into main text Eq.~6
yields
\begin{equation}
    \Gcwb[(t)] = \frac{\sum
        \left(\exppathlength[\alpha,*]{i}{j}(0)-\exppathlength[\alpha,*]{i}{j}\left(\left\{S_{ab}\right\}\right)\right)}
    {\sum\left(\exppathlength[\alpha,*]{i}{j}(0)-\exppathlength[\alpha,*]{i}{j}(1)\right)}
    = \frac{\sum
        \left(\alpha \cdot \exppathlength[1,*]{i}{j} - \exppathlength[\alpha,*]{i}{j}\left(\left\{S_{ab}\right\}\right)\right)}
    {\sum\left(\alpha \cdot \exppathlength[1,*]{i}{j} - \exppathlength[1,*]{i}{j}\right)}
    = \frac{\sum
        \left(\alpha \cdot \exppathlength[1,*]{i}{j} - \exppathlength[\alpha,*]{i}{j}\left(\left\{S_{ab}\right\}\right)\right)}
    {\left(\alpha - 1\right)\cdot \sum \exppathlength[1,*]{i}{j}}
\end{equation}
where for better legibility, we omitted the summation indices over all reachable destinations.
Expanding the sum in the numerator results in
\begin{equation}
    \Gcwb[(t)] = \frac{1}{\alpha - 1} \left[\alpha -
        \frac{\sum\limits_{i \in \Vsrc{}, j \in \Vdst{}}\exppathlength[\alpha,*]{i}{j}\left(\left\{S_{ab}\right\}\right)}
        {\sum\limits_{i \in \Vsrc, j \in \Vdst{}}\exppathlength[1,*]{i}{j}}\right]
    \label{eq:coolwalkability_simplified}
\end{equation}
which is still correct for general street networks. Each vertex in a grid is
identified by two coordinates $\left(x,y\right)$ pointing at a column and row,
relative to an arbitrarily chosen origin $\left( 0,0 \right)$. A path is then
denoted by $\left( x_1,y_1 \right) \rightarrow \left( x_2, y_2 \right)$. Using
the translational invariance of the infinite grid, any such path is equivalent to the
same path shifted to the origin,
\begin{equation}
    \left( x_1,y_1 \right) \rightarrow \left( x_2, y_2 \right)
    \equiv
    \left( 0,0 \right) \rightarrow \left( x_2-x_1, y_2-y_1 \right).
\end{equation}
As such, the sums over index $i$ in equation \ref{eq:coolwalkability_simplified} only result in a 
factor
$\abs{\Vsrc{}}$ which cancels out. For only the expression in
question we get

\begin{equation}
    \frac{\sum\limits_{i \in \Vsrc{}, j \in \Vdst{}}\exppathlength[\alpha,*]{i}{j}\left(\left\{S_{ab}\right\}\right)}
    {\sum\limits_{i \in \Vsrc{}, j \in \Vdst{}}\exppathlength[1,*]{i}{j}}
    =
    \frac{\sum\limits_{\tup{x}{y} \in \Vdst[(\tup{0}{0})]{}}\exppathlength[\alpha,*]{\tup{0}{0}}{\tup{x}{y}}\left(\left\{S_{ab}\right\}\right)}
    {\sum\limits_{\tup{x}{y} \in \Vdst[(\tup{0}{0})]{}}\exppathlength[1,*]{\tup{0}{0}}{\tup{x}{y}}} \label{eq:coolwalkability_grid_step1}
\end{equation}
On a rectangular grid with two types of edges $a$ and $b$ the length of any shortest path is given by
\begin{equation}
    \exppathlength[\alpha,*]{\tup{x_1}{y_1}}{{\tup{x_2}{y_2}}} = \abs{x_2 - x_1} \cdot \lf[a]{} +\abs{y_2 - y_1} \cdot \lf[b]{} = \abs{\Delta x} \cdot \lf[a]{} + \abs{\Delta y} \cdot \lf[b]{}
\end{equation}
which, in Eq.~\ref{eq:coolwalkability_grid_step1} gives
\begin{equation}
    \frac{\sum\limits_{\tup{x}{y} \in \Vdst[(\tup{0}{0})]{}}\exppathlength[\alpha,*]{\tup{0}{0}}{\tup{x}{y}}\left(\left\{S_{ab}\right\}\right)}
    {\sum\limits_{\tup{x}{y} \in \Vdst[(\tup{0}{0})]{}}\exppathlength[1,*]{\tup{0}{0}}{\tup{x}{y}}} =
    \frac{
        \sum\limits_{\tup{\Delta x}{\Delta y} \in \Vdst[]{}}
        \abs{\Delta x} \cdot \lf[a]{} + \abs{\Delta y} \cdot \lf[b]{}
    }
    {\sum\limits_{\tup{\Delta x}{\Delta y} \in \Vdst[]{}}
        \abs{\Delta x} \cdot \ls[a]{} + \abs{\Delta y} \cdot \ls[b]{}
    }
\end{equation}
with
\begin{equation}
    n=\sum\limits_{(\Delta x, \Delta y) \in \Vdst[]{}}\abs{\Delta x} \quad\quad\quad m = \sum\limits_{(\Delta x, \Delta y) \in \Vdst[]}\abs{\Delta y}
\end{equation}
we get the \Cwb{} \ref{eq:coolwalkability_simplified} on the grid as
\begin{align}
\Gcwb[(t)] &= \frac{1}{\alpha-1} \left[\alpha - \frac{n \cdot \lf[a]{} + m \cdot \lf[b]{}}{n \cdot \ls[a] + m \cdot \ls[b]}\right] \notag\\
 &= \frac{1}{\alpha-1} \left[\alpha - \frac{n \cdot \left(\alpha \cdot \lh[a] + \lc[a]\right) + m \cdot \left(\alpha \cdot \lh[b] + \lc[b]\right)}{n \cdot \ls[a] + m \cdot \ls[b]}\right] \notag\\
 &= \frac{1}{\alpha-1} \left[\alpha - \frac{n \cdot \left(\alpha \left[\ls[a] - \lc[a]\right] + \lc[a]\right) + m \cdot \left(\alpha\left[\ls[b]- \lc[b]\right]+\lc[b]\right)}{n \cdot \ls[a] + m \cdot \ls[b]}\right] \notag\\
 &= \frac{1}{\alpha-1} \left[\alpha - \frac{\alpha \left[n \, \ls[a] + m \, \ls[b]\right] + \left(1-\alpha\right)\left[n \, \lc[a]+ m \, \lc[b] \right]}{n \cdot \ls[a] + m \cdot \ls[b]}\right]
\end{align}
Simplifying this equation gives the shape of the resulting main text Eq.~7:
\begin{equation}
    \Gcwb = \frac{1}{\alpha-1} \left[\alpha - \alpha - \left(1-\alpha\right) \frac{n\,\lc[a]+m\,\lc[b]}{n\,\ls[a] + m\,\ls[b]}\right] = \frac{n\,\lc[a]+m\,\lc[b]}{n\,\ls[a] + m\,\ls[b]} \label{eq:coolwalkability_grid_shape}
\end{equation}

\subsection*{CoolWalkability in the large trip-length limit}
To solve the equation
\begin{equation}
  n =\sum\limits_{(\Delta x, \Delta y) \in \Vdst[]}\abs{\Delta x}
  \label{eq:solving_n_start_appendix}
\end{equation}
we need to describe the set \Vdst[], which contains all the vertices
of the grid reachable within $r$ in Manhattan distance from the center.
Geometrically, it contains all grid-point within a square with edge length of
$\sqrt{2}\,r$ centered around $(0,0)$ with the diagonals aligned with the
directions of the grid. The maximal number of jumps in $x$ direction is therefore
\begin{equation}
  \Delta x_{\max} = \left\lfloor\frac{r}{\ls[a]}\right\rfloor
\end{equation}
given a number of jumps $\Delta x$ in $x$ direction, we find the maximal number
of jumps possible in $y$ direction as
\begin{equation}
  \Delta y_{\max}(\Delta x) = \left\lfloor\frac{r-\ls[a]\abs{\Delta x}}{\ls[b]}\right\rfloor
\end{equation}
using the symmetry of the grid we express equation
\ref{eq:solving_n_start_appendix} as a double sum with dependent limits
\begin{equation}
  n =\sum\limits_{(\Delta x, \Delta y) \in \Vdst[]}\abs{\Delta x} = \sum_{\substack{\Delta x = \\-\Delta x_{\max}}}^{\Delta x_{\max}}\left(\sum_{\substack{\Delta y =\\-\Delta y_{\max}(\Delta x)}}^{\Delta y_{\max}(\Delta x)} \abs{\Delta x}\right)
\end{equation}
where we directly evaluate the inner sum as
\begin{equation}
  n = \sum_{\substack{\Delta x = \\-\Delta x_{\max}}}^{\Delta x_{\max}}\abs{\Delta x}\cdot\left(2\Delta y_{\max}(\Delta x)+1\right)
  \label{eq:n_exact_final}
\end{equation}
due to the gauss-brackets in $\Delta x_{\max}$ and more importantly $\Delta
  y_{\max}(\Delta x)$, we can not simplify this expression any further, but solving
it numerically is very much possible.
By expressing the summation limits in terms of $\Delta y$
\begin{align}
  \Delta y_{\max} = \left\lfloor\frac{r}{\ls[b]}\right\rfloor \quad\quad\quad \Delta x_{\max}(\Delta y) = \left\lfloor\frac{r-\ls[b]\abs{\Delta y}}{\ls[a]}\right\rfloor
\end{align}
we find a similar expression for the value of $m$
\begin{equation}
  m = \sum_{\substack{\Delta y = \\-\Delta y_{\max}}}^{\Delta y_{\max}}\abs{\Delta y}\cdot\left(2\Delta x_{\max}(\Delta y)+1\right)
  \label{eq:m_exact_final}
\end{equation}
Assuming a large radius compared to the lengths of the individual edges, we
replace the double summation with a double integral
\begin{equation}
  n =\sum_{\substack{\Delta x = \\-\Delta x_{\max}}}^{\Delta x_{\max}} \sum_{\substack{\Delta y =\\-\Delta y_{\max}(\Delta x)}}^{\Delta y_{\max}(\Delta x)} \abs{\Delta x}
  \quad\rightarrow\quad
  \int\limits_{-\frac{r}{\ls[a]}}^{\frac{r}{\ls[a]}} \int\limits_{-\frac{r-\ls[a]\abs{x}}{\ls[b]}}^{\frac{r-\ls[a]\abs{x}}{\ls[b]}} \abs{x} \,\mathrm{d}y\,\mathrm{d}x =
  \frac{4}{\ls[b]\left(\ls[a]\right)^2}\int\limits_{0}^{r} \int\limits_{0}^{r-z} z \,\mathrm{d}y\,\mathrm{d}z
\end{equation}
where the integral evaluates to
\begin{equation}
  \int\limits_{0}^{r} \int\limits_{0}^{r-z} z \,\mathrm{d}y\,\mathrm{d}z =
  \int\limits_{0}^{r} rz - z^2 \,\mathrm{d}z = \frac{r^3}{2} - \frac{r^3}{3} = \frac{1}{6}r^3
\end{equation}
such that we get
\begin{equation}
  n = \frac{2}{3}\,\frac{r^3}{\ls[b]\left(\ls[a]\right)^2}
\end{equation}
and, by analogous calculations,
\begin{equation}
  m = \frac{2}{3}\,\frac{r^3}{\ls[a]\left(\ls[b]\right)^2}
\end{equation}
Finally, plugging these results into equation \ref{eq:coolwalkability_grid_shape} yields main text Eq.~7:
\begin{equation}
    \frac{n\,\lc[a]+m\,\lc[b]}{n\,\ls[a] + m\,\ls[b]} = \frac{\frac{2}{3}\,\frac{r^3}{\ls[b]\left(\ls[a]\right)^2}\,\lc[a]+\frac{2}{3}\,\frac{r^3}{\ls[a]\left(\ls[b]\right)^2}\,\lc[b]}{\frac{2}{3}\,\frac{r^3}{\ls[b]\left(\ls[a]\right)^2}\,\ls[a] + \frac{2}{3}\,\frac{r^3}{\ls[a]\left(\ls[b]\right)^2}\,\ls[b]} =
     \frac{\ls[b]\,\lc[a]+\ls[a]\,\lc[b]}{2\ls[a]\ls[b]}
\end{equation}

\section*{Supplementary Note 3: Extending the range}\label{appendix:extensions}
It is important to extend the range of \SI{800}{\metre} to check how longer distances affect \Cwb{}. We indeed managed to triple the range from \SI{800}{\metre} to \SI{2400}{\metre} on the centerline networks. With up to \SI{2400}{\metre} walk lengths, we can be confident that we cover distances that pedestrians would actually walk on a hot day, for example between their home and place of work or between touristic points of interest -- if one wanted to focus on individual pedestrians. See the last columns in the new Figs.~SI7-12. 

The increased range has minor expected effects on \Cwb{}, such as ``smoothening'' the curves, as larger distances cause each starting node to reach more parts of the city, possibly traversing multiple, structurally different local neighborhoods, which causes the individual signatures of these neighborhoods to blur into each other. For example, in Manhattan (Fig.~SI7) the diurnal ``W'' profile becomes less pronounced when extending the maximum walking distance in the centerline network (compare panel \textbf{B} with \textbf{D}) as many of the trips are now able to extend beyond the downtown area of Manhattan with high buildings and a clearly oriented grid, into neighborhoods of the city where the buildings are not as tall and the street grid is oriented differently. 

The extended range covers more green areas and thus also has the natural effect of increasing differences between the scenarios that neglect and that incorporate parks, compare panels \textbf{I} and \textbf{K} in Figs.~SI7-9. The biggest difference is in Manhattan (Fig.~SI7) due to the inclusion of central park.

Due to our city-focused approach, where we assess CoolWalkability for each point in the city as an aggregate of potential walks from that point to all points in its local neighborhood, it is already appropriate to deal with distances that are not too long (i.e., \SI{800}{\metre}) to quantify the \emph{local} structures of the city. Nevertheless, the extension to \SI{2400}{\metre} provides an important robustness check.

\section*{Supplementary Note 4: Realized routes and impacts}\label{appendix:realizedroutes}
Within the context of our model, a hypothetical user would input their personal sun aversion $\alpha$ as the answer to the question: ``How much farther are you willing to walk in the shade compared to in the sun?'' Our model then finds the shortest experienced route, which is, by this definition, optimal. However, this process does not quantify how much better, if at all, the optimal route is, compared to the physically shortest route. To clarify this question, we study two simple measures. 

The first measure is the relative distance traveled in the sun when using shaded routes, compared to the distance traveled in the sun when using the physically shortest paths $\frac{\Lh{\alpha}}{\Lh{1}}$ defined in main text Eq.~10, where $\Lh{\alpha}$ is the total length traveled in the sun when optimizing for shaded routes with a sun aversion of $\alpha$. The results are reported in panels E-H of Figs.~SI10-12. Optimizing for shaded routes causes the relative distance in the sun during the day to decrease to values mostly between \SI{20}{\percent} and \SI{90}{\percent}, depending on the city, time of day, and sun aversion. For a given route, these values might help an individual decide whether it deems the proposed shaded route beneficial, compared to the shortest path. However, a small value does not necessarily imply a good performance of a given city, as the structure of the city and the available shade might limit the effect of shaded routing: For example, during noon, when there is little shade available ($S\approx0$), the physically shortest paths are nearly the same as the experienced shortest paths, simply due to a lack of alternatives, which results in a high relative distance. To the contrary, during mornings and evenings, when the city is fully shaded ($S\approx1$), the physically shortest paths are already efficient, and shaded routing might not affect these routes, leading again to a high value. See for example, Fig.~SI10E, where $\frac{\Lh{\alpha}}{\Lh{1}}$ is close to 1 for low $\alpha$ while \Cwb{} is also close to 1. In this case, the total distances in the sun are generally fairly short, and thus even small routing changes can have strong effects, causing the resulting relative distance in the sun to fluctuate, explaining the large range of values for different $\alpha$ values. This becomes especially apparent at the end of the day (mornings/evenings). 

The second measure is the relative increase in physical trip-length when optimizing for shaded routes compared to the physical length of the physically shortest paths $\frac{\Ltot{\alpha}}{\Ltot{1}}$ defined in main text Eq.~11, where $\Ltot{\alpha}$ is the total length traveled in the sun when optimizing for shaded routes with a sun aversion of $\alpha$. The results are reported in panels G-I of Figs.~SI10-12. The overall physical length of all paths increases only by up to \SI{30}{\percent} at $\alpha=10$ and stays below \SI{10}{\percent} for values of $\alpha\leq2$. These values are all considerably lower than the theoretically possible increase by a factor of $\alpha$, as it does not happen in practice that all shortest paths are fully covered in sun while all experienced shortest paths are fully covered in shade. Concerning the concrete shapes of the daily $\frac{\Ltot{\alpha}}{\Ltot{1}}$ curves, it is hard to find general patterns. In general, the curves rise in the morning and fall in the evening, showing different shapes in between. Sometimes, we observe that the curves dip or fluctuate between 12:00 and 16:00. This fluctuation might be due to a ``breakdown of alternatives'': as the amount of available shade decreases as the sun rises, longer and longer detours are necessary to find the shortest experienced path, especially for high $\alpha$. However, at a certain point, the available shade is either not sufficient or not distributed in such a way as to facilitate shaded walks. At this point, the physically shortest paths (or some paths close to them) again become favorable, and the relative physical distance collapses. This observation, however, makes the relative distance-increase unsuitable as a measure for the performance of a city, as we do not know whether a low value is due to a favorable shade distribution or due to a general lack of options.

To understand the positive effects of reduced distance traveled in the sun, it might be beneficial to express the reduction in terms of the avoided heat-stress as well as UV exposure. While a detailed discussion of the health implications goes beyond our model, both these effects are roughly proportional to the distance walked in the sun, and as such are, in a first approximation, described by the relative distance traveled in the sun.

\clearpage
\section*{Supplementary Figures}

\begin{figure*}[h]
    \centering
    \includegraphics[width=0.9\textwidth]{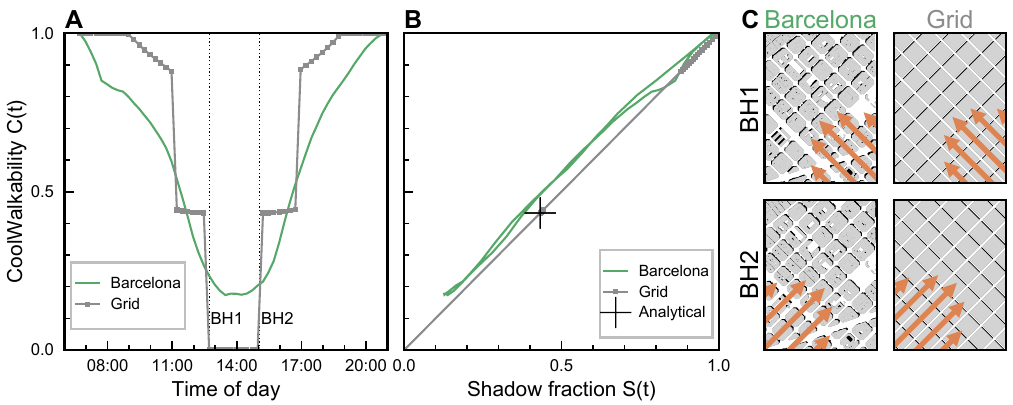}
    \caption{\textbf{Diurnal \Cwb{} profile and phase portrait for Barcelona.} This is a companion figure to main text Fig.~3. \textbf{A:} Due to lower, more uniform building heights, there are no noticeable ``Barcelonahenge'' dips (BH1 and BH2) as in Manhattan. \textbf{B:} The phase portrait shows little positive deviation of the empirical data from the grid model, implying only little \Cwb{} benefits at a given shadow fraction. \textbf{C:} Sun position showing the closest Barcelona and its grid model get to ``henge'' events.}
    \label{fig:barcelona_overview_supplement}
\end{figure*}

\begin{figure*}[h]
    \centering
    \includegraphics[width=0.9\textwidth]{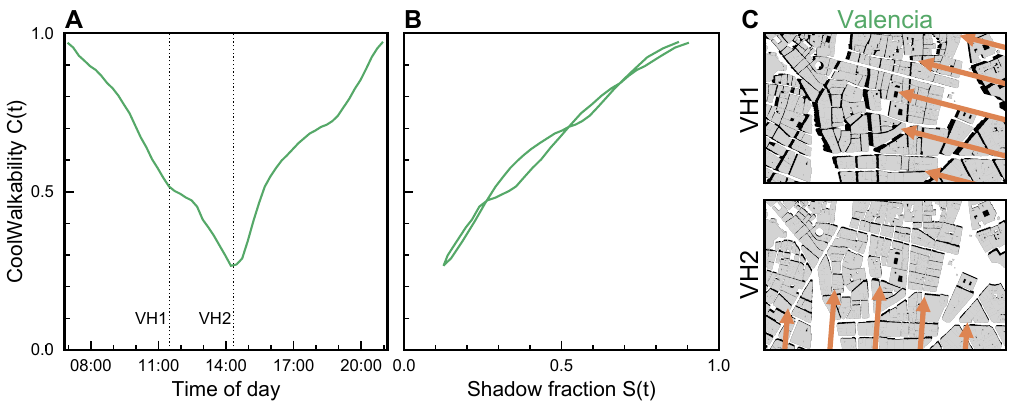}
    \caption{\textbf{Diurnal \Cwb{} profile and phase portrait for Valencia.} This is a companion figure to main text Fig.~3. \textbf{A:} Due to a non-grid-like street network and lower, more uniform building heights, there are no noticeable ``Valenciahenge'' events (VH1 and VH2) as in Manhattan. \textbf{B:} The phase portrait is similar to Barcelona, Fig.~SI1, implying only little \Cwb{} benefits at a given shadow fraction. \textbf{C:} Sun position showing the closest Valencia gets to ``henge'' events.}
    \label{fig:valencia_overview_supplement}
\end{figure*}

\begin{figure*}[h]
    \centering
    \includegraphics[width=0.7\textwidth]{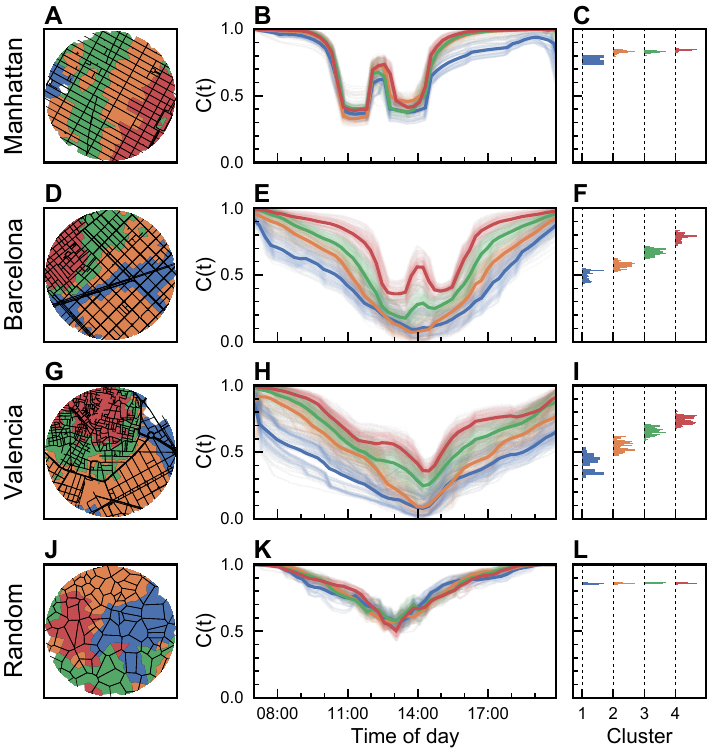}
    \caption{\textbf{Spatial clustering by CoolWalkability leads to areas with different profiles (constant building heights).} This is a companion figure to main text Fig.~5, reporting the same results but for constant building heights. From top to bottom, we study the cities Manhattan, Barcelona, Valencia, and the random null model (Poisson-Voronoi). Left column: Clustering local Coolwalkability of each node in the street network leads to spatial clusters of similar CoolWalk potential. Middle column: The diurnal profiles of these clusters display high variations within each city and between different cities. In particular, the more organic, least grid-like areas (red curves) display highest potential. \textbf{K:} The null model shows the baseline of small variation. Right column: the distributions of the time average of each diurnal profile within each cluster illustrate the large potential differences in empirical street networks. \textbf{L:} These differences are negligible in the null model.}
    \label{si:fig:clustering}
\end{figure*}

\begin{figure*}[h]
    \centering
    \includegraphics[width=0.7\textwidth]{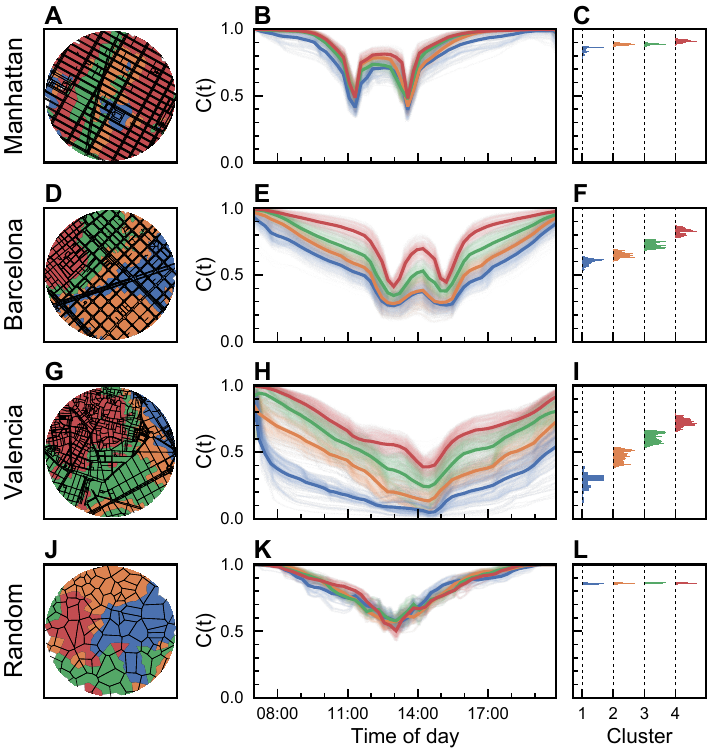}
    \caption{\textbf{Spatial clustering by CoolWalkability leads to areas with different profiles (sidewalk network).} This is a companion figure to main text Fig.~5, reporting the same results but for the sidewalk networks. From top to bottom, we study the cities Manhattan, Barcelona, Valencia, and the random null model (Poisson-Voronoi). Left column: Clustering local Coolwalkability of each node in the street network leads to spatial clusters of similar CoolWalk potential. Middle column: The diurnal profiles of these clusters display high variations within each city and between different cities. In particular, the more organic, least grid-like areas (red curves) display highest potential. \textbf{K:} The null model shows the baseline of small variation. Right column: the distributions of the time average of each diurnal profile within each cluster illustrate the large potential differences in empirical street networks. \textbf{L:} These differences are negligible in the null model.}
    \label{si:fig:clustering}
\end{figure*}

\begin{figure*}[h]
    \centering
    \includegraphics[width=0.7\textwidth]{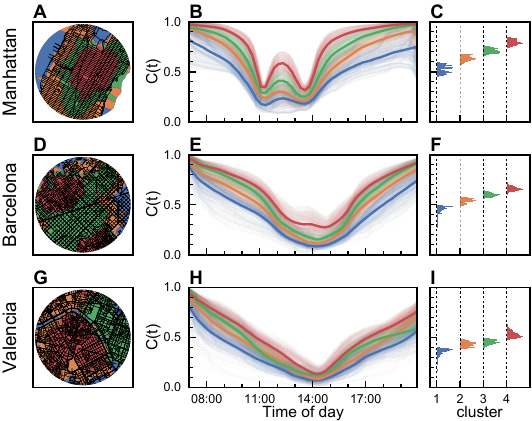}
    \caption{\textbf{Spatial clustering by CoolWalkability leads to areas with different profiles (centerline network, \SI{2400}{\metre}).} This is a companion figure to main text Fig.~5, reporting the same results but for larger maximal walking distances. From top to bottom, we study the cities Manhattan, Barcelona and Valencia. Left column: Clustering local Coolwalkability of each node in the street network leads to spatial clusters of similar CoolWalk potential. Middle column: The diurnal profiles of these clusters display high variations within each city and between different cities. In particular, the more organic, least grid-like areas (red curves) display highest potential.}
    \label{si:fig:clustering}
\end{figure*}

\begin{figure*}[h]
    \centering
    \includegraphics[width=0.7\textwidth]{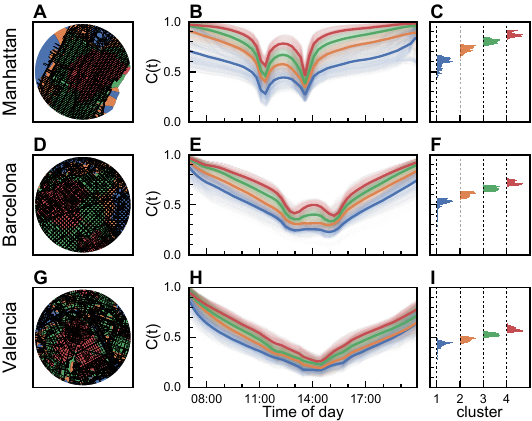}
    \caption{\textbf{Spatial clustering by CoolWalkability leads to areas with different profiles (full network, \SI{2400}{\metre}).} This is a companion figure to main text Fig.~5, reporting the same results but for larger maximal walking distances. From top to bottom, we study the cities Manhattan, Barcelona and Valencia. Left column: Clustering local Coolwalkability of each node in the street network leads to spatial clusters of similar CoolWalk potential. Middle column: The diurnal profiles of these clusters display high variations within each city and between different cities. In particular, the more organic, least grid-like areas (red curves) display highest potential.}
    \label{si:fig:clustering}
\end{figure*}

\begin{figure*}[h]
    \centering
    \includegraphics[width=0.7\textwidth]{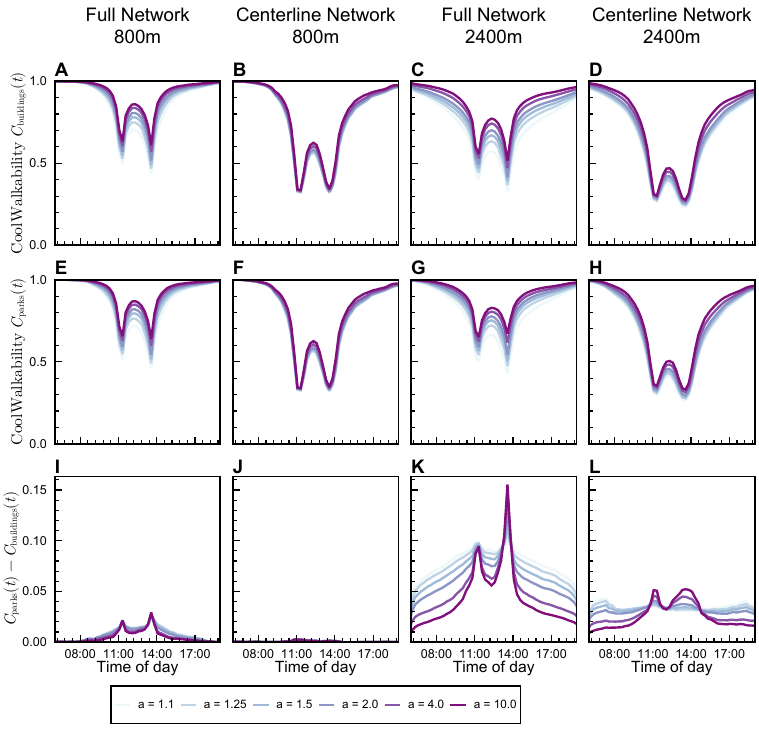}
    \caption{\textbf{The small effect of parks on the global \Cwb{} (Manhattan)} This figure extends the results reported in the main text by testing larger maximal trip lengths, as well as the inclusion of parks. From top to bottom, we show \textbf{A-D} the diurnal \Cwb{}-profiles for the city as defined in the main text, \textbf{E-H} for the city with the inclusion of parks, and \textbf{I-L} the difference in \Cwb{} between a city with and without parks. The difference is strongest during noon and the Manhattan-Henge events, where trees cast shade on the paths below them, while buildings provide only little shade to the paths next to them. In particular, it is strongest for the full network with extended walking distances (panel \textbf{K}), as more of the trips can be rerouted through central park.
    }
    \label{si:fig:parks_manhattan}
\end{figure*}

\begin{figure*}[h]
    \centering
    \includegraphics[width=0.7\textwidth]{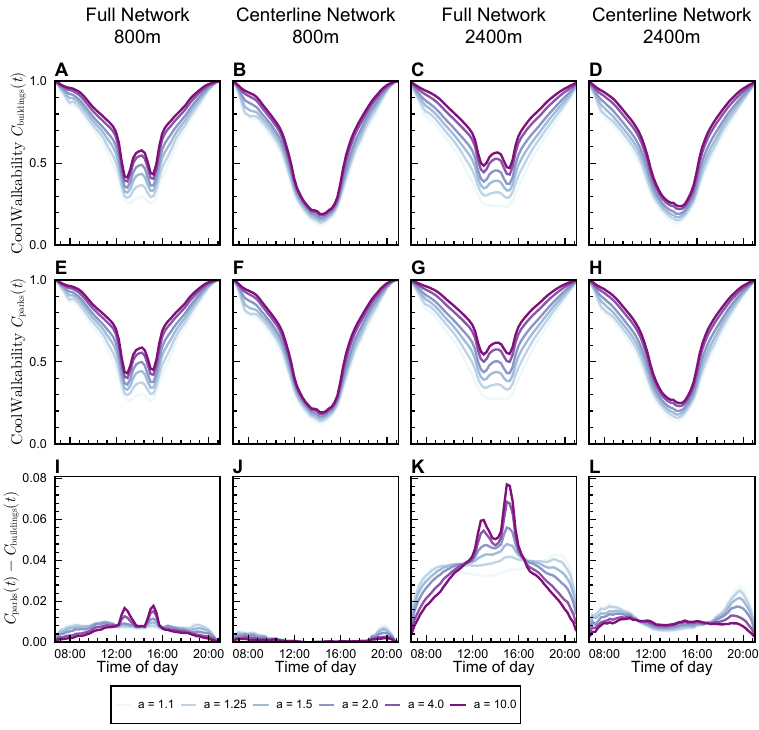}
    \caption{\textbf{The small effect of parks on the global \Cwb{} (Barcelona)} This figure extends the results reported in the main text by testing larger maximal trip lengths, as well as the inclusion of parks. From top to bottom, we show \textbf{A-D} the diurnal \Cwb{}-profiles for the city as defined in the main text, \textbf{E-H} for the city with the inclusion of parks, and \textbf{I-L} the difference in \Cwb{} between a city with and without parks.
    }
    \label{si:fig:parks_barcelona}
\end{figure*}

\begin{figure*}[h]
    \centering
    \includegraphics[width=0.7\textwidth]{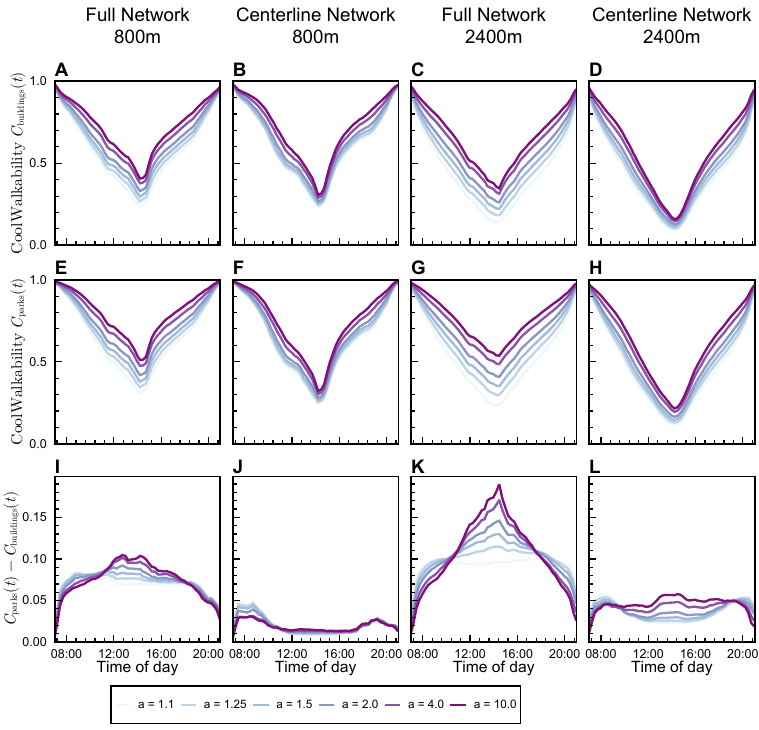}
    \caption{\textbf{The small effect of parks on the global \Cwb{} (Valencia)} This figure extends the results reported in the main text by testing larger maximal trip lengths, as well as the inclusion of parks. From top to bottom, we show \textbf{A-D} the diurnal \Cwb{}-profiles for the city as defined in the main text, \textbf{E-H} for the city with the inclusion of parks, and \textbf{I-L} the difference in \Cwb{} between a city with and without parks.}
    \label{si:fig:parks_valencia}
\end{figure*}

\begin{figure*}[h]
    \centering
    \includegraphics[width=0.7\textwidth]{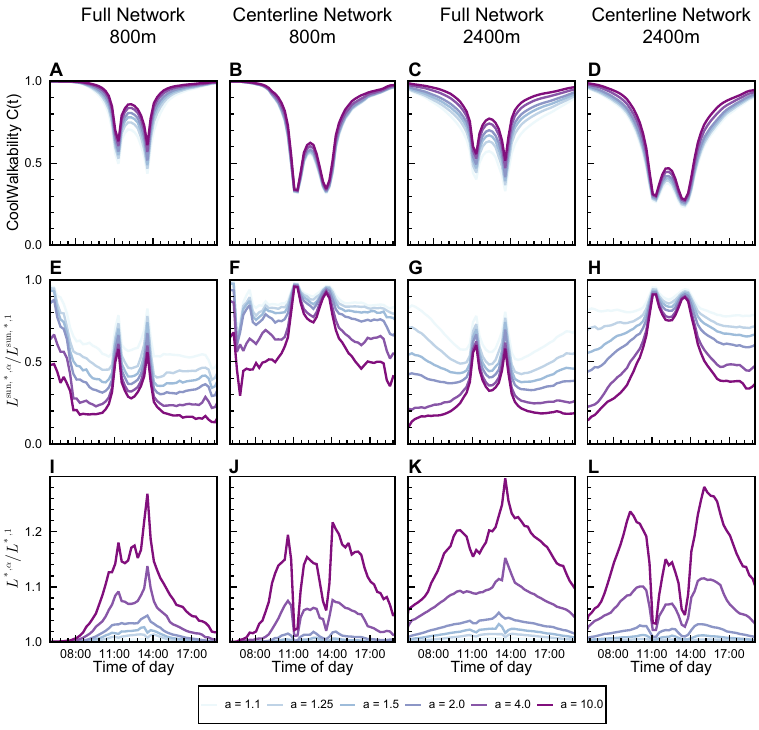}
    \caption{\textbf{Realized impacts compared to \Cwb{} (Manhattan)} This figure compares alternative measure based on the overall physical distances traveled in the sun and shade to the \Cwb{}. From top to bottom, we show \textbf{A-D} the diurnal \Cwb{}-profiles for the city as defined in the main text, \textbf{E-H} the relative distance traveled in the sun on all trips between using shaded routes (at $\alpha > 1$) and the physically shortest paths (at $\alpha=1$) as well as \textbf{I-L} the relative total physical distance between $\alpha>1$ and $\alpha =1$. The relative distance in the sun generally decreases with increasing values of $\alpha$, and increases, during noon.}
    \label{si:fig:real_lengths_manhattan}
\end{figure*}

\begin{figure*}[h]
    \centering
    \includegraphics[width=0.7\textwidth]{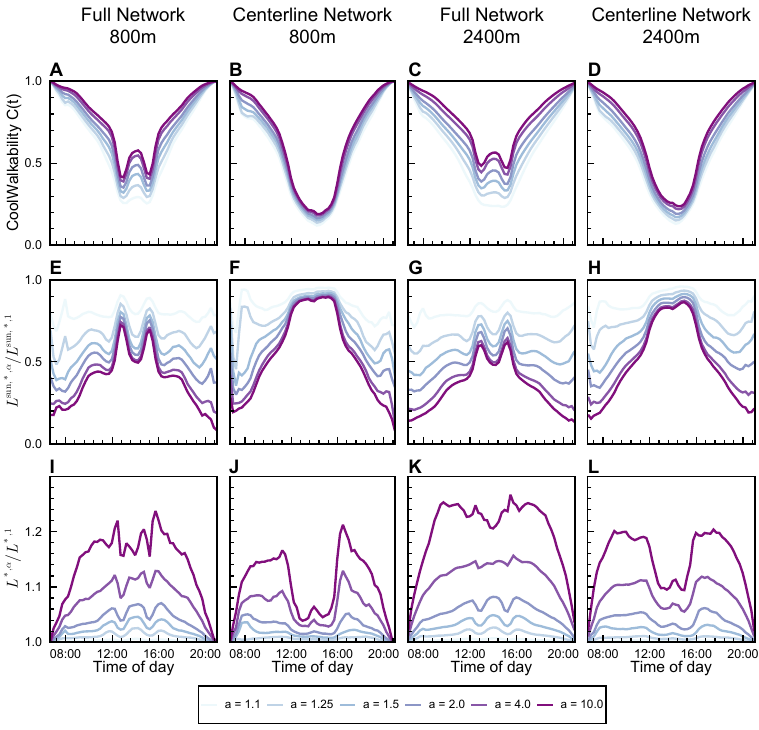}
    \caption{\textbf{Realized impacts compared to \Cwb{} (Barcelona)} This figure compares alternative measure based on the overall physical distances traveled in the sun and shade to the \Cwb{}. From top to bottom, we show \textbf{A-D} the diurnal \Cwb{}-profiles for the city as defined in the main text, \textbf{E-H} the relative distance traveled in the sun on all trips between using shaded routes (at $\alpha > 1$) and the physically shortest paths (at $\alpha=1$) as well as \textbf{I-L} the relative total physical distance between $\alpha>1$ and $\alpha =1$. The relative distance in the sun generally decreases with increasing values of $\alpha$, and increases, during noon.}
    \label{si:fig:real_lengths_barcelona}
\end{figure*}

\begin{figure*}[h]
    \centering
    \includegraphics[width=0.7\textwidth]{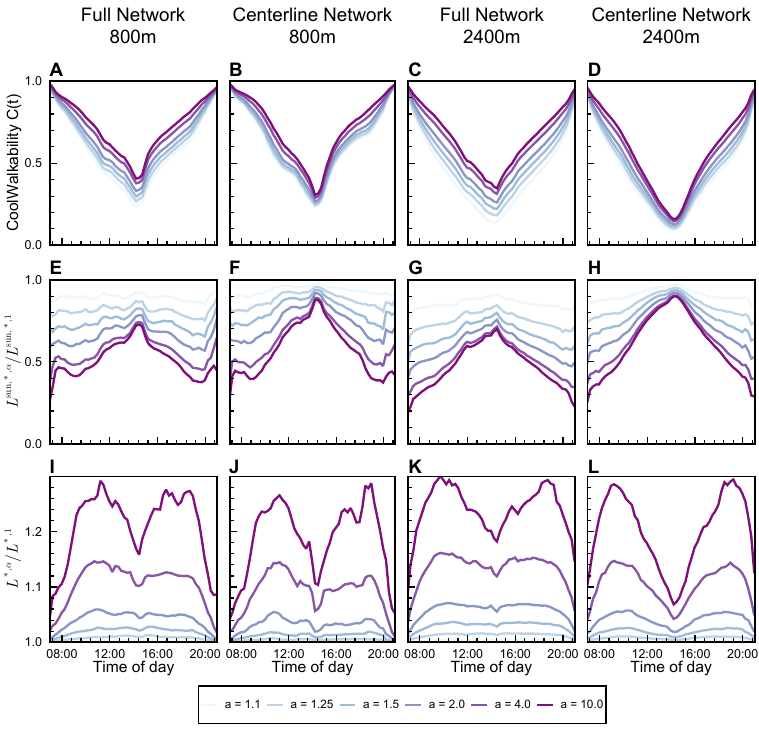}
    \caption{\textbf{Realized impacts compared to \Cwb{} (Valencia)} This figure compares alternative measure based on the overall physical distances traveled in the sun and shade to the \Cwb{}. From top to bottom, we show \textbf{A-D} the diurnal \Cwb{}-profiles for the city as defined in the main text, \textbf{E-H} the relative distance traveled in the sun on all trips between using shaded routes (at $\alpha > 1$) and the physically shortest paths (at $\alpha=1$) as well as \textbf{I-L} the relative total physical distance between $\alpha>1$ and $\alpha =1$. The relative distance in the sun generally decreases with increasing values of $\alpha$, and increases, during noon.}
    \label{si:fig:real_lengths_valencia}
\end{figure*}

\begin{figure*}[h]
    \centering
    \includegraphics[width=0.7\textwidth]{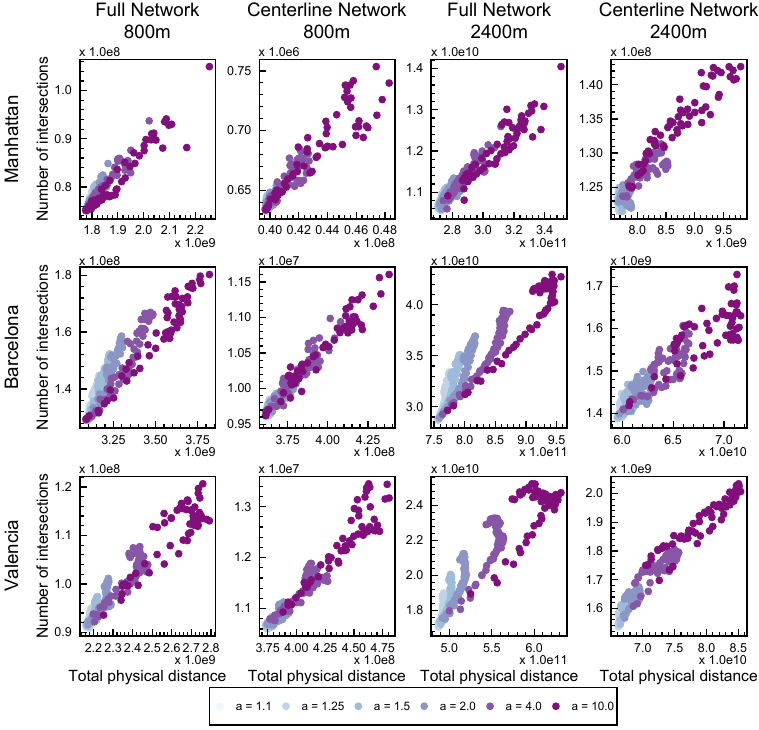}
    \caption{\textbf{Number of intersections encountered compared to the total distance traveled} Increasing $\alpha$ potentially increases the total physical length of the trips traveled. With this increase in physical length, pedestrians tend to encounter more intersections, increasing the perceived complexity of a route. This figure shows that the number of intersections encountered is roughly linear in the total physical distance, showing that shade-aware routing does not increase the complexity of the experienced shortest paths beyond the complexity added due to an increased length. Only in the full network, particularly at \SI{2400}{\metre} in Barcelona and Valencia we observe some non-linear increase for individual $\alpha$ values. In these particular cases it might happen that more paths can be routed through both highly shaded and intersection-dense neighborhoods of the city, thus increasing the overall number of encountered intersections beyond the linear regime.}
    \label{si:fig:num_intersections_small}
\end{figure*}

\begin{figure*}[h]
    \centering
    \includegraphics[width=0.7\textwidth]{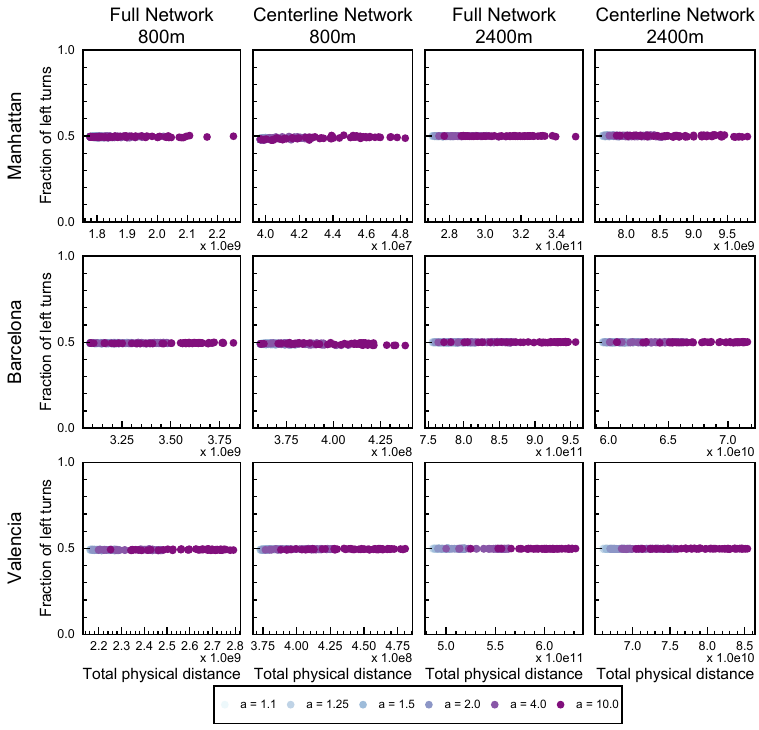}
    \caption{\textbf{Fraction of left turns compared to the total distance traveled} Counting the number of turns to the left of the direction of travel and comparing this number to the full number of turns, we find no systematic bias towards either left or right turns in any of the studied cities with a change in $\alpha$ or the time of day.}
    \label{si:fig:left_turns_small}
\end{figure*}